\documentclass[11pt]{article} 
\usepackage[a4paper, margin=1.25in]{geometry} 
\usepackage{graphicx}
\usepackage{lipsum}

\usepackage[dvipsnames,svgnames,x11names]{xcolor}

\usepackage{balance}
\usepackage[T1]{fontenc}
\usepackage[normalem]{ulem}

\usepackage[numbers]{natbib}

\usepackage{array}
\newcolumntype{L}[1]{>{\raggedright\let\newline\\\arraybackslash\hspace{0pt}}m{#1}}
\newcolumntype{C}[1]{>{\centering\let\newline\\\arraybackslash\hspace{0pt}}m{#1}}
\newcolumntype{R}[1]{>{\raggedleft\let\newline\\\arraybackslash\hspace{0pt}}m{#1}}

\usepackage{listings}

\lstdefinelanguage{XML}
{
basicstyle=\ttfamily\footnotesize,
  morestring=[b]",
  moredelim=[s][\bfseries\color{Maroon}]{<}{\ },
  moredelim=[s][\bfseries\color{Maroon}]{</}{>},
  moredelim=[l][\bfseries\color{Maroon}]{/>},
  moredelim=[l][\bfseries\color{Maroon}]{>},
  morecomment=[s]{<?}{?>},
  morecomment=[s]{<!--}{-->},
  commentstyle=\color{gray},
  stringstyle=\color{blue},
  identifierstyle=\color{red}
}

\usepackage{moreverb}

\usepackage[nounderscore]{syntax}

\DeclareGraphicsExtensions{.pdf}

\usepackage[cmex10]{amsmath}
\usepackage{amssymb}
\usepackage{mathtools}
\usepackage{amsthm}
\usepackage{amsfonts}

\usepackage{subfig} 
\usepackage{makecell}

\usepackage{algorithmicx}
\usepackage{algpseudocode}
\usepackage[ruled]{algorithm}
\definecolor{light-gray}{gray}{0.75}
\algrenewcommand{\algorithmiccomment}[1]{\hskip3em{{\footnotesize \textcolor{light-gray}{$\blacktriangleright$}}} #1}

\usepackage{multirow} 
\usepackage{rotating} 
\usepackage{booktabs} 
\usepackage{colortbl} 
\usepackage{tablefootnote}

\usepackage[pdftex,colorlinks=true,urlcolor=blue,citecolor=blue]{hyperref}

\usepackage{xspace}

\usepackage{enumitem}

\usepackage{blindtext}

\usepackage{listings}
\usepackage{xcolor}

\definecolor{codegreen}{rgb}{0,0.6,0}
\definecolor{codegray}{rgb}{0.5,0.5,0.5}
\definecolor{codepurple}{rgb}{0.58,0,0.82}
\definecolor{backcolour}{rgb}{0.95,0.95,0.92}

\lstdefinestyle{mystyle}{
    backgroundcolor=\color{backcolour},   
    commentstyle=\color{codegreen},
    keywordstyle=\color{magenta},
    numberstyle=\tiny\color{codegray},
    stringstyle=\color{codepurple},
    basicstyle=\ttfamily\footnotesize,
    breakatwhitespace=false,         
    breaklines=true,                 
    captionpos=b,                    
    keepspaces=true,                 
    numbers=left,                    
    numbersep=5pt,                  
    showspaces=false,                
    showstringspaces=false,
    showtabs=false,                  
    tabsize=2,
    escapeinside={(*@}{@*)}
}

\newcommand{\ad}{AeroDaaS\xspace}
\newcommand{\adp}{AeroDaaS\xspace}

\begin{document}

\title{\huge \adp: A Programmable Drones-as-a-Service Platform for Intelligent Aerial Systems}

\author{
Kautuk Astu, Suman Raj, Priyanshu Pansari and Yogesh Simmhan\\~\\
Department of Computational and Data Sciences,\\
Indian Institute of Science, Bangalore 560012 India\\~\\
Email: \{kautukastu, sumanraj, simmhan\}@iisc.ac.in
}

\date{}

\maketitle

\begin{abstract}
The increasing adoption of UAVs equipped with advanced sensors and GPU-accelerated edge computing has enabled real-time AI-driven applications in domains such as precision agriculture, wildfire monitoring, and environmental conservation. 
However, the integrated design and orchestration of navigation, sensing, and analytics, together with seamless real-time coordination across drone, edge, and cloud resources, remains a significant challenge.
To address these challenges, we propose \textit{\adp}, a service-oriented framework that abstracts UAV-based sensing complexities and provides a \textit{Drone-as-a-Service} (DaaS) model for intelligent decision-making. \adp offers modular service primitives for on-demand UAV sensing, navigation and analytics as composable microservices, ensuring cross-platform compatibility and scalability across heterogeneous UAV and edge-cloud infrastructures. \adp also supports plug-and-play scheduling modules, including \textit{Waypoint} and \textit{Analytics} schedulers, which enable trajectory optimization and real-time coordination of inference workloads.
We implement and evaluate \adp for six real-world DaaS applications, of which two are evaluated in real-world scenarios and four in simulation.
\adp requires $\leq40$ lines of code for the applications and has minimal platform overhead of $\leq20$ ms per frame and $\leq1$ GB memory usage on Orin Nano and a AMD RTX 3090 GPU workstation. These  results are promising for \adp as an efficient, flexible and scalable UAV programming framework for autonomous aerial analytics.
\end{abstract}

\section{Introduction}
Unmanned Aerial Vehicles (UAVs), commonly referred to as drones~\footnote{We use UAVs and drones interchangeably in this article.}, equipped with onboard computing and communication capabilities are increasingly deployed as on-demand aerial services for applications such as disaster response~\cite{liu2023dome,10556878,KUMAR2024108977},
precision agriculture~\cite{QU2024108543,betti2024drone},
urban traffic management~\cite{10556942,BARMPOUNAKIS202050}, and accessibility~\cite{raj2023ocularone,avila2017dronenavigator}. They rely on deep learning-driven analytics, such as object detection and anomaly identification, using real-time camera feeds and sensor data for autonomous decision-making. 
\textit{Drones-as-a-Service (DaaS)} extends this paradigm by enabling UAVs to be accessed as scalable, autonomous services for diverse domains. 
DaaS has been proposed for logistics, where drones optimize package delivery while reducing environmental impact~\cite{shahzaad2019composing}, and in healthcare, for timely transport of medical supplies to disaster-affected areas~\cite{alkouz2020swarm,shahzaad2020game}.

Advances in edge computing and networked architectures
further enhance DaaS by enabling distributed processing and multi-drone coordination~\cite{chu2021holistic,li2024uav}. 
For example, in an urban safety scenario, a UAV that is observing various regions of a city may detect a specific stolen car or missing person, and start following it, and once emergency responders intercept the entity, the UAV resumes its broader mission.
However, these require custom application design, joint scheduling of trajectory and analytics~\cite{10381761,khochareTON} and runtime optimizations for edge-cloud offloading and drone coordination~\cite{10108067}.

\subsection{Challenges}

Companies like DJI, Parrot, and Skydio offer diverse drone platforms, requiring \textit{proprietary} Software Development Kits (SDKs) such as DJI Mobile SDK~\cite{dji_mobile_sdk} to program them. These offer APIs for \textit{low-level control} over drone navigation and on-board sensors. Custom drone applications developed using these SDKs are limited to that specific vendor, and often support a specific drone model as the SDK capabilities differ.
This prevents their \textit{portability} to other drones and substantially increases developer effort.
Further, these SDKs \textit{lack high-level primitives} for task-driven application composition. Practical DaaS applications require real-time \textit{analytics} for autonomous navigation and intelligence. Given the pervasiveness of Deep Neural Network (DNN) models over vision and sensor data, these analytics also require \textit{edge accelerators}, such as Nvidia Jetson, present onboard or co-located with the drone for rapid inferencing~\cite{uav_edgeinf,10171496}.
These must be complemented by cloud resources and Inferencing-as-a-Service, intermittently accessible through cellular networks~\cite{8943331,RAJ2025107874,he2021confect}. 

This also requires intelligent scheduling of analytics and offloading from on-board edge devices to the cloud~\cite{raj2024adaptiveheuristicsschedulingdnn}. 
Further, when analytics drive navigation decisions, such as dynamically prioritizing triggered events over routine missions~\cite{10.1145/3386901.3388912}, the challenge extends to \textit{co-scheduling analytics and drone navigation}. This close-coupling of analytics with trajectory scheduling highlights the need for unified frameworks that can jointly manage analytics, task priorities, and navigation in DaaS frameworks.

Lastly, given the safety-critical nature of drone applications, they must also be tested in \textit{simulation environments} like \textit{Gazebo} before deployment~\cite{1389727}. This requires the same DaaS application written for physical drones and environments to also support different virtual drone platforms and scenarios, and also linking the simulation environment with the analytics runtime on edge and cloud. This will reduce the friction between development, testing and reliable deployment.

\subsection{Gaps}
The proprietary SDKs offered by commercial drone manufacturers are incompatible with others. While they offer niche consumer-facing features such as $360^{\circ}$ rotation to capture selfies, they have limited higher-order primitive to ease programming~\cite{dji_mobile_sdk}.
The open-source drone ecosystem has autopilot frameworks like PX4~\cite{px4_web_ref} and ArduPilot~\cite{ardupilot_web_ref} with extensible interfaces for UAV navigation. 
However, they operate at a low level abstraction, with SDKs limited to controller-level commands. They are also incompatible with proprietary drone platforms and lack high-level primitives for analytics and edge-cloud operations.

AeroStack2~\cite{fernandez2023aerostack2} and UAL~\cite{real2020unmanned} offer hardware-level abstractions for cross-platform drone interoperability, functioning primarily as ``device drivers''. While we build upon AeroStack2, these frameworks lack higher-level primitives for analytics and orchestration. Others introduce Domain-Specific Languages (DSLs) for targeted tasks such as video sensing~\cite{10.1145/3486607.3486750}, mission optimization~\cite{10.1145/3386901.3388912} or swarm coordination~\cite{10.1109/IROS.2016.7759558} but do not generalize to other drone applications. 
However, these approaches fall short in offering a unified programming framework that can seamlessly compose sensing, navigation and analytics, and schedule the analytics across edge and cloud resources. This highlights the critical need for intuitive higher-order primitives and analytics orchestration to design and deploy DaaS applications for heterogeneous drone and computing platforms.

\subsection{Contributions}
In this article, we introduce \textbf{\adp}, a programmable DaaS platform that advances interoperable  and seamless composition of analytics-driven drone applications across the edge-cloud continuum alongside modular sensing and  trajectory planning for diverse operational scenarios.

The key contributions of this paper are:
\begin{enumerate}[leftmargin=*]
\item We introduce the \textit{\adp architecture} and define a \textit{DaaS Programming Framework} motivated by real-world DaaS scenarios that integrates sensing, navigation and analytics to concisely compose DaaS applications (\S \ref{sec:prog-model}). 
\item We extend this with modular analytics and navigation schedulers from literature across edge and cloud to optimally perform analytical tasks and generate waypoints.
\item We describe an initial \textit{runtime implementation} of \adp that supports multiple drone platforms, including Gazebo simulator (\S \ref{sec:runtime}).
\item We perform a detailed evaluation of \adp for VIP Tracking with situation awareness in the real-world using physical drones, and Smart crop monitoring in a simulation environment to demonstrate the flexibility of our model and validate the edge-cloud analytics (\S~\ref{sec:evals}).
\end{enumerate}
Besides these, we review related work (\S \ref{sec:related}), offer background (\S \ref{sec:bg}), and discuss conclusions and future work (\S \ref{sec:conclusions}).

An earlier short version~\cite{aerodaas} of this article motivated the need for programming abstractions in DaaS frameworks, introduced the \ad architecture to support sensing, navigation, and analytics for composing DaaS applications, provided APIs to integrate DNN analytics across the edge-cloud continuum, and demonstrated its use for a basic navigation task in simulation and a complex analytics task on real hardware. The current article extends it substantially by first discussing various navigation patterns (\S~\ref{sec:bg}) to motivate priority-aware trajectory planning, and then enhancing the APIs by introducing \textit{INavigate} and updating other interfaces as well as the architecture. Next, \adp defines workflows to realize the navigation patterns (\S~\ref{sec:workflows}). Finally, we evaluate these contributions across an extensive set of scenarios, both in simulation using Gazebo and on hardware using Jetson Orin Nano and a Ryze Tello drone in outdoor campus environments (\S~\ref{sec:evals}).
Our prior poster paper~\cite{10898177} demonstrates integration of DNN analytics but does not exploit the edge-cloud continuum or complex compositions, which are addressed by \ad~\cite{aerodaas}.

\section{Related Work}
\label{sec:related}

\begin{table*}[t]
\centering
\caption{Comparison of Service-Oriented Frameworks for Drones-as-a-Service (DaaS) Applications}
\label{tab:related-work}
\scriptsize
\setlength{\tabcolsep}{1pt}
\begin{tabular}{c||C{2.5cm}|C{2cm}|c|c|c|C{2cm}|C{1.5cm}}
\toprule
\textbf{Framework} & \textbf{Programming Interface} & \textbf{Cross-Hardware} & \textbf{Navigation} & \textbf{Sensing} & \textbf{Analytics} & \textbf{General\-iza\-bility} & \textbf{Edge + Cloud}  \\ \midrule
\textbf{UAL}   \cite{real2020unmanned}      & Python          & \checkmark & \checkmark & \checkmark & \texttimes & \checkmark & \texttimes  \\
\textbf{Aerostack2} \cite{fernandez2023aerostack2}  & Python          & \checkmark & \checkmark & \checkmark & \texttimes & \checkmark & \texttimes  \\
\textbf{SoftwarePilot} \cite{10.1145/3565386.3565484}  & Java          & \texttimes & \checkmark & \checkmark & \checkmark & \texttimes & \texttimes  \\
\textbf{SkyQuery} \cite{10.1145/3486607.3486750}    & Custom DSL      & \checkmark & \texttimes & \texttimes & \checkmark & \texttimes & \texttimes  \\
\textbf{Buzz} \cite{10.1109/IROS.2016.7759558}        & Buzz language   & \texttimes & \checkmark & \checkmark & \texttimes & \checkmark & \texttimes  \\
\textbf{VOLTRON} \cite{10.1145/2668332.2668353}    & Java and C++    & \checkmark & \texttimes & \checkmark & \texttimes & \checkmark & \texttimes  \\
\textbf{PaROS} \cite{10.1145/3197768.3197772}     & Java            & \checkmark & \checkmark & \checkmark & \texttimes & \checkmark & \texttimes  \\
\textbf{BeeCluster} \cite{10.1145/3386901.3388912} & Python          & \texttimes & \texttimes & \checkmark & \texttimes & \checkmark & \texttimes  \\
\textbf{AnDrone} \cite{van2019androne}    & Android Things  & \texttimes & \checkmark & \checkmark & \texttimes & \checkmark & \texttimes  \\ \midrule
\rowcolor[HTML]{DFFFD6}
\textbf{AeroDaaS}    & Python   & \checkmark & \checkmark & \checkmark & \checkmark & \checkmark & \checkmark \\ \bottomrule
\end{tabular}
\end{table*}

This section reviews existing frameworks, tools, and methodologies for UAV services, analyzing their contributions and limitations.

\subsection{Cross-Platform UAV Programming}

Several frameworks provide low-level hardware abstractions for drones, focusing on control and actuation rather than higher-level services. AeroStack2~\cite{fernandez2023aerostack2} offers a ROS2-based abstraction layer, managing UAV hardware through a plugin-oriented architecture.  AeroDaaS builds on AeroStack2, automating infrastructure generation while extending capabilities to analytics integration, orchestration, and multi-device collaboration. UAV Abstraction Layer (UAL)~\cite{real2020unmanned} defines basic UAV functions but lacks extensibility and cross-platform compatibility. AeroDaaS provides a higher-level service layer, enabling analytics-driven applications and distributed coordination beyond UAL’s foundational controls. SoftwarePilot 2.0~\cite{10.1145/3565386.3565484} employs a microservices-based approach, containerizing mission logic using Docker and Kubernetes. However, its complex deployment and DJI-only support limit usability. \adp simplifies deployment while ensuring scalability, interoperability, and service-driven orchestration across diverse drone platforms.

\subsection{Domain Specific Languages (DSL) for UAVs}
Several frameworks leverage Domain-Specific Languages (DSLs) to abstract low-level drone operations. SkyQuery\cite{10.1145/3486607.3486750} enables video analytics and waypoint generation but lacks direct drone control services. Buzz\cite{10.1109/IROS.2016.7759558} provides a swarm programming DSL, but its reliance on the Buzz Virtual Machine (BVM) and absence of built-in analytics primitives add complexity. BeeCluster~\cite{10.1145/3386901.3388912} optimizes multi-drone missions using predictive analytics, yet lacks cross-platform compatibility. 
\adp goes beyond DSL limitations by providing a service-driven, platform-agnostic framework that integrates analytics, navigation, and multi-device orchestration, enabling seamless UAV application development without low-level complexities.

\subsection{Programming Framework for UAVs}

VOLTRON\cite{10.1145/2668332.2668353} simplifies multi-drone coordination but is restricted to sensing services. PaROS\cite{10.1145/3197768.3197772} offers swarm orchestration primitives, automating specific missions but lacking DNN-based analytics services that leverage edge accelerators. AnDrone~\cite{van2019androne} virtualizes drones using Linux containers called Android Things. Users can configure virtual drones in the cloud, utilizing existing Android apps and resources, which are then safely deployed on real drone hardware. AnDrone is limited to Android apps, and offers just navigation and sensing capabilities. We use a similar Docker-based approach that can work across the edge and cloud continuum. 

\subsection{Software Development Kits (SDKs) for UAVs}
Most commercial SDKs, such as the DJI SDKs~\cite{dji_mobile_sdk}, Parrot GroundSDK~\cite{parrot_sdk}, and Autel SDK~\cite{autel_sdk}, are proprietary and tightly coupled to vendor-specific hardware, leading to vendor lock-in and poor portability. Open-source frameworks like PX4~\cite{px4_web_ref}, ArduPilot~\cite{ardupilot_web_ref}, and Paparazzi UAV~\cite{paparazzi_uav} offer greater flexibility but mainly expose low-level flight control and telemetry APIs without unified abstractions for sensing, analytics, or orchestration. Middleware such as MAVSDK~\cite{mavsdk}, DroneKit~\cite{dronekit}, and AirSim~\cite{airsim} support cross-platform communication and simulation but require extensive customization for multi-UAV or edge–cloud integration. Due to their reliance on MAVLink-based control, DroneKit and AirSim have already announced their deprecation. Lightweight APIs like the Crazyflie Python API~\cite{crazyflie} and Dronecode SDK–Python~\cite{dronecode_sdk_python} are similarly limited in scalability. Finally, MATLAB’s UAV and Robotics System Toolkit~\cite{matlab_uav} enables high-level modeling and simulation but remains confined to the proprietary MATLAB/Simulink ecosystem and is heavily reliant on PX4 for interoperability with real-world experiments. Unlike these SDKs, which are primarily developed for roboticists and require an understanding of low-level expertise, \adp exposes high-level, declarative APIs that allow users to have a smoother learning curve.

\subsection{Integration of Task and Route Schedulers with DaaS}
The integration of task scheduling and route planning in DaaS frameworks has become a critical aspect in aerial edge computing. Several studies have focused on either optimizing the drone trajectories or improving computational task allocation, but real-world conditions may require scheduling in both aspects. For instance, Hamadi~\cite{9380171} and Alkouz~\cite{9872142} investigated the drone service selection problem under varying drone behaviors and service models, yet they assumed static flight plans and ignored the arrival of new tasks during operation. Similarly, Wang \textit{et al.}~\cite{11023224} proposed a task offloading mechanism for UAVs but did not account for spatial locality among drones or adaptive trajectory optimization. Gao \textit{et al.}~\cite{10994279} explored a joint trajectory–resource optimization scheme, but its applicability is limited to lightweight tasks that UAVs can execute while flying, without considering dynamically evolving tasks, which may have an impact on the trajectory. Similarly, the Serv-HU framework~\cite{10814719} achieves smooth task migration between MEC servers by determining optimal spatio-temporal handover points; however, it ignores unforeseen trajectory deviations. In contrast, \adp provides a modular architecture that seamlessly integrates both the task scheduler and the route scheduler as plugins, enabling flexible and co-optimized coordination between computational and mobility decisions.

\subsection{Contrast with Prior Work}

\adp significantly advances over \ad~\cite{aerodaas} by expanding both the architectural and experimental scope. It first formalizes a taxonomy of UAV navigation patterns, which motivates the design of priority-aware trajectory planning. The core architecture is further refined with the addition of \textit{INavigate} abstraction and enhancements to other APIs, enabling extensible integration of both task and route scheduling logic. Building upon these extensions, the proposed \adp framework illustrates executable workflows that capture diverse navigation and scheduling patterns. Finally, \adp provides a comprehensive evaluation across Gazebo simulations to real-world experiments on Jetson Orin Nano and DJI Ryze Tello drones in outdoor campus deployments.

\section{Motivation and DaaS Requirements}
\label{sec:bg}

\begin{figure*}[!t]
    \centering
    \subfloat[Pre-defined mission with static waypoints]{
    \includegraphics[width=0.35\textwidth]{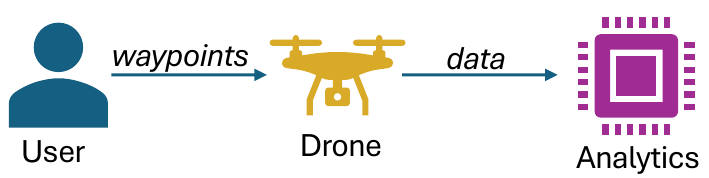}
    \label{fig:short-preplanned}
    }
    \subfloat[Analytics-driven mission abort]{
    \includegraphics[width=0.35\textwidth]{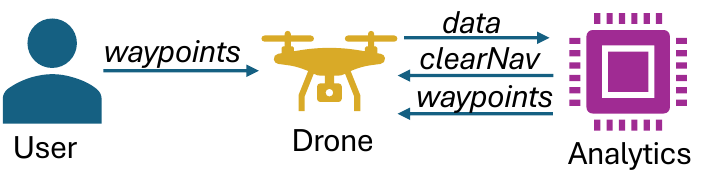}
    \label{fig:short-mission-abort}
    }
    \subfloat[Sensor data-driven mission
    ]{
    \includegraphics[width=0.3\textwidth]{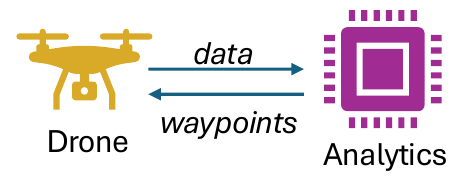}
    \label{fig:short-sensor-driven}
    }\\
    \subfloat[Analytics-driven mission update
    ]{
    \includegraphics[width=0.4\textwidth]{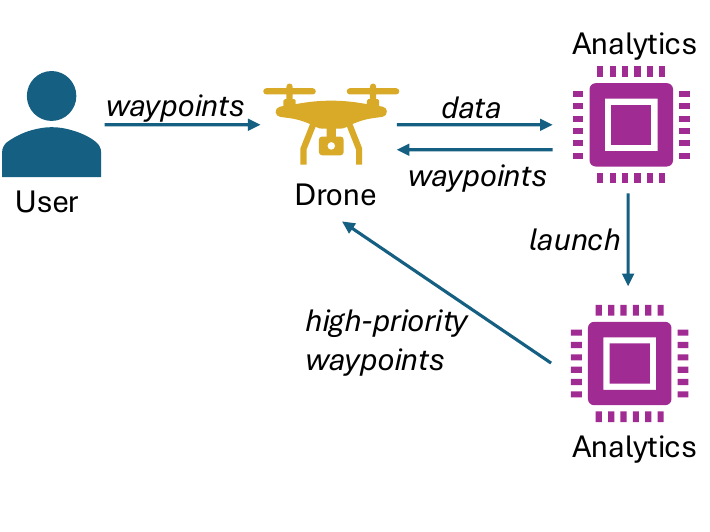}
    \label{fig:short-mission-change}
    }\qquad
    \subfloat[Pre-defined mission with dynamic waypoints
    ]{
    \includegraphics[width=0.4\textwidth]{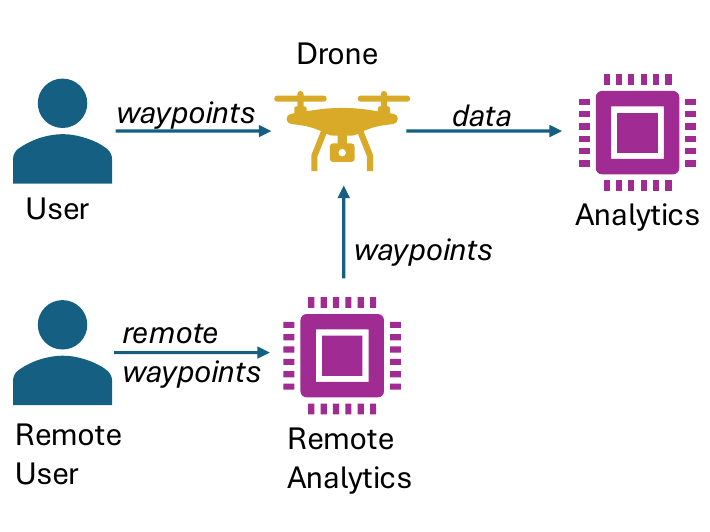}
    \label{fig:short-remote}
    }
\caption{Different patterns of interaction between User-Analytics-Drone}
\label{fig:patterns}
\end{figure*}

In this section, we discuss various mission patterns motivated by real-world drone scenarios, and subsequently outline the requirements these patterns impose on \adp for their effective composition.

\subsection{Drone Service Paradigms}

We identify two representative drone service paradigms that leverage a \textit{sensing-analytics-navigation} loop for a DaaS programming framework: (1)~\textit{Analytics-Driven Services (ADS)}: Drones autonomously adjust their trajectory based on real-time analytics. E.g., UAV-based assistive services can track a visually impaired person by detecting their hazard vest using a DNN, issue navigation commands, and provide obstacle alerts~\cite{raj2023ocularone,RAJ2025107874}. In addition to assistive services, UAVs can track moving vehicles to improve traffic management and monitor critical infrastructures~\cite{10318456}. (2)~\textit{Waypoint-Driven Services (WDS)}: Drones follow a predefined trajectory based on waypoints, and execute sensing or analytics tasks at specific locations during the flight. E.g., UAVs capture traffic images along city roads to assess congestion and optimize traffic signaling~\cite{khan2017uav,khochareTON}, or be used in farm surveying~\cite{betti2024drone}.

\subsection{Taxonomy of Mission Patterns}
\label{subsec:mission-pattern}
Next, we discuss $5$ major patterns observed in drone missions. The simplest one is (a) \textit{Pre-defined mission with static waypoints} as shown in Fig.~\ref{fig:short-preplanned}. Here, the drone follows a set of waypoints predefined by the user, which remain unchanged during the mission. While the drone is in flight, an onboard analytics can be optionally run based on the collected data. A complex version of this is (b) \textit{Pre-defined mission with dynamic waypoints} as shown in Fig.~\ref{fig:short-remote}. Here, the drone initially follows a set of predefined waypoints, but new waypoints can be added to the set during the mission. These can be updated either through a remote user or an independent analytics server, allowing for dynamic adjustments based on observations or changing mission requirements. Both of these patterns follow the WDS paradigm. Another trivial pattern motivated by autonomous navigation is (c) \textit{Sensor data-driven mission} in which the navigation of the drone is controlled solely by the commands generated using real-time onboard analytics of the live drone feeds from onboard camera (Fig.~\ref{fig:short-sensor-driven}). This leverages the ADS paradigm.

We also discuss two complex hybrid patterns that combine characteristics of ADS and WDS. In (d) \textit{Analytics-driven mission update}, the drone’s mission initially consists of predefined user waypoints; however, onboard analytics can generate high-priority waypoints (triggers) during flight. This enables the drone to dynamically switch from its ongoing mission to address a high-priority task based on real-time analysis or triggered events, and subsequently resume its original mission (Fig.~\ref{fig:short-mission-change}). A related variant is shown in (e) \textit{Analytics-driven mission abort}, where the onboard analytics have the authority to completely override the current mission. Upon detecting or triggering a critical event, the analytics module clears the existing navigation queue and generates new high-priority waypoints, preventing the drone from returning to its prior mission as in the previous pattern (Fig.~\ref{fig:short-mission-abort}).

\subsection{Requirements}

Based on these mission patterns, we highlight requirements of \adp, which we address in this article.

\begin{enumerate}[leftmargin=*]
   \item \textit{Composable Interfaces for Common Application Logic}: 
A unified framework with APIs for \textit{sensing, navigation and analytics} is essential for addressing common drone application needs. While existing solutions provide limited support based on specific drone hardware, they lack cross-platform compatibility. E.g., DJI SDKs offer mission-specific services such as \textit{Revolve}, \textit{Follow}, and waypoint-based navigation, but these do not extend to other drones.

\item 
\textit{Hardware-Agnostic Implementation}:
A unified, service-oriented stack that modularizes sensing, navigation, and analytics is required to enable seamless deployment across diverse drone hardware. Simulation tools should be further supported to test without altering application logic.

\item 
\textit{Service Deployment and Execution Isolation}: To ensure seamless deployment of drone applications with diverse dependencies across heterogeneous compute environments, we require a containerized service model that encapsulates applications and inferencing services in isolated environments, enabling conflict-free execution, improved security, and efficient resource utilization.

\item 
\textit{Analytics Services across Edge-Cloud Continuum}: Given the computational demands of DNN-based analytics, resource constraints on the mission computers (edge devices) can limit performance. Applications must define priorities and performance guarantees for analytics tasks, necessitating seamless integration of intelligent analytics schedulers that manage DNN inferencing services across the edge-cloud continuum to meet latency and throughput requirements. 

\item 
\textit{Intelligent Trajectory Scheduling:} To ensure efficient and reliable operation of drone within the onboard battery constraints, there is a need of trajectory scheduler which optimizes drone trajectory, minimises battery usage, reduces service time of waypoints, maximizes throughput by serving maximum number of waypoints. Applications should define the coordinates, priorities, and deadlines for the waypoints, enabling seamless integration with the trajectory scheduler. The scheduler should be able to efficiently handle waypoints from both Analytics-Driven Services and Waypoint-Driven Services.

\end{enumerate}

\section{Architecture and Programming Framework}\label{sec:prog-model}

\begin{figure}[t]
\centering
\includegraphics[width=0.6\columnwidth]{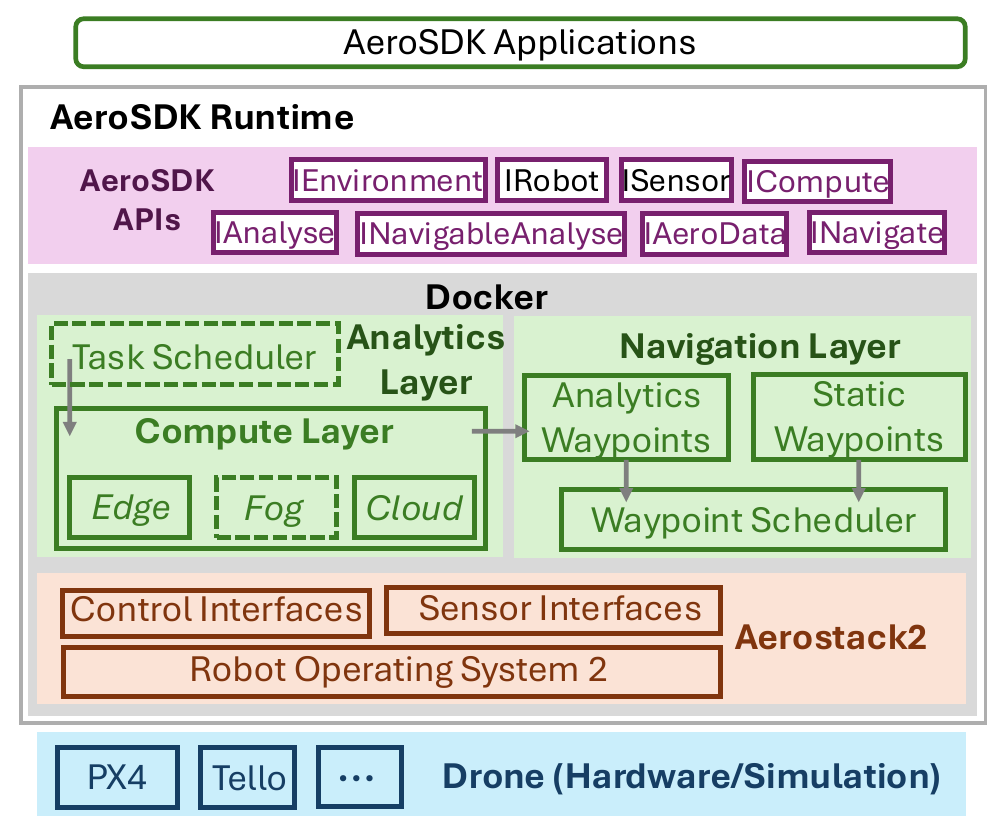}
\caption{\adp Architecture and Services} 
\label{fig:architecture}
\end{figure}

\adp offers core programming APIs and data models to enable developers to intuitively compose DaaS applications. 
It builds on core components of the drone ecosystem by abstracting their complexity into modular services. It leverages AeroStack2 as the hardware abstraction layer for seamless integration with Pixhawk~\cite{5980229} as the flight controller. It offloads compute-intensive tasks such as DNN inferencing to companion mission computers like the Nvidia Jetson or remote cloud resources. By extending autopilot platforms such as PX4~\cite{px4_web_ref} and ArduPilot~\cite{ardupilot_web_ref}, \adp enables analytics-driven autonomous navigation. Also, it utilizes offboard APIs in high-level languages like Python to bridge drones with external compute and sensing resources, ensuring flexible, programmable, and scalable control across edge and cloud.

\subsection{\adp Architecture}
\label{sec:arch}
We describe the architecture of \adp using Fig.~\ref{fig:architecture}.
DaaS applications are built using \adp's APIs (Table.~\ref{tab:aerodaas-api}, which abstracts low-level drone navigation, controls, sensing and analytics into intuitive, composable Python-based interfaces. Internally, some of these APIs extend Aerostack2's capabilities by integrating with onboard sensors and actuators across both real-world and simulated environments.

The key novelty in \adp lies in its Analytics and Navigation  service layers (shown in green). The analytics layer enables DNN-based models to run across edge, fog and cloud resources for a hybrid execution model. These analytics services can be intelligently used by plugging in task schedulers to dynamically distribute workloads to balance performance and cost.
The navigation layer, on the other hand, incorporates a waypoint scheduler that optimizes UAV trajectories to achieve an optimal mission completion based on user requirements. Moreover, this layer provides a modular interface that allows seamless integration of diverse path-planning algorithms, facilitating adaptability across application domains and deployment environments.
To ensure modularity and isolation, \adp employs containerized service execution using Docker. Each service, such as sensor access, analytics and compute management, runs in an independent container, allowing seamless deployment and encapsulation between layers.

\begin{table*}[t!]
\centering
\caption{APIs provided by \adp}
\label{tab:aerodaas-api}
\scriptsize
\setlength{\tabcolsep}{1pt}
\begin{tabular}{C{2cm}||C{2.25cm}|C{2.25cm}|C{2.25cm}||L{7cm}}
\hline
\textbf{Interface} & \textbf{Function} & \textbf{Input} & \textbf{Output} & \textbf{Description} \\ 
\hline \hline

\multirow{9}{*}{IEnvironment} 
& hasEnvSensors & -- & boolean & Returns True if the environment has sensors. \\ 
\noalign{\global\arrayrulewidth=0.1pt}\arrayrulecolor{lightgray} \cline{2-5}
& getEnvSensors & -- & List<ISensor> & Returns a list of sensors in the environment. \\ \cline{2-5}
& getEnvSensorByID & str & ISensor & Returns the requested sensor to the user. \\ \cline{2-5}
& getEnvRobots & -- & List<IRobot> & Returns a list of all the robots available in the environment. \\ \cline{2-5}
& getRobotByID & str & IRobot & Returns the requested robot to the user. \\ \cline{2-5}
& isRobotSensorAvailable & str & boolean & Returns True if a robot has required onboard sensor available. \\ \cline{2-5}
& getRobotSensors & -- & List<ISensor> & Returns a list of all the sensors present on all robots in environment. \\ \cline{2-5}
& getComputeResources & -- & List<ICompute> & Returns all the compute resources available. \\ \cline{2-5}
& getComputeResourceByID & str & ICompute & Returns the requested compute resource to the user. \\ \cline{2-5}
& setTrajectoryScheduler & INavigate & -- & Plugs in the trajectory scheduler in the environment. \\ \cline{2-5}
& setAnalyticsScheduler & str & -- & Plugs in the analytics scheduler in the environment.\\ 
\noalign{\global\arrayrulewidth=0.4pt}\arrayrulecolor{black} \hline

\multirow{9}{*}{IRobot} 
& startMission & -- & -- & Hook for the user code to be run before starting a mission. \\ 
\noalign{\global\arrayrulewidth=0.1pt}\arrayrulecolor{lightgray}
\cline{2-5}
& endMission & -- & -- & Hook for the user code to be run after a mission ends. \\ \cline{2-5}
& pauseMission & -- & -- & Pause the ongoing mission by putting the drone in hover mode. \\ \cline{2-5}
& resumeMission & -- & -- & Resume the ongoing mission. \\ \cline{2-5}
& hasSensors & -- & boolean & Returns True if the robot has any onboard sensors. \\ \cline{2-5}
& getSensors & -- & List<ISensor> & Returns a list of all the sensors present on board. \\ \cline{2-5}
& getSensorByID & str & ISensor & Returns the requested sensor. \\ \cline{2-5}
& isNavigable & -- & boolean & Returns True if a robot is navigable. \\ \cline{2-5}
& navigate & IAeroData<AeroNav> & -- & Invokes navigation commands based on navigation data. \\ 
\noalign{\global\arrayrulewidth=0.4pt}\arrayrulecolor{black}
\hline

\multirow{3}{*}{ISensor}
& getSensorProperty & str & str & Get properties of a sensor. \\ 
\noalign{\global\arrayrulewidth=0.1pt}\arrayrulecolor{lightgray}
\cline{2-5}
& getSensorID & -- & str & Returns the unique ID of a sensor. \\ \cline{2-5}
& getDataStream & -- & AeroStreamData & 
Returns an iterator implementing IAeroData to yield sensor data
\\
\noalign{\global\arrayrulewidth=0.4pt}\arrayrulecolor{black} \hline

\multirow{2}{*}{IAeroData}
& getData & -- & E & Returns a data item of type E. \\ 
\noalign{\global\arrayrulewidth=0.1pt}\arrayrulecolor{lightgray}
\cline{2-5}
& setData & E & -- & Sets a data item of type E. \\ 
\noalign{\global\arrayrulewidth=0.4pt}\arrayrulecolor{black}
\hline
ICompute & getComputeProperties & -- & dict & Returns the properties of the compute resource. \\ \hline
& analyse & IAeroData & IAeroData & Invokes an analysis function. \\ 
\noalign{\global\arrayrulewidth=0.1pt}\arrayrulecolor{lightgray}
\cline{2-5}
IAnalyse & instantiate & -- & IAeroData & Invokes an analytic function which receives data from external sources. \\ \cline{2-5}
& deploy & ICompute & -- & Assigns the compute device for executing analysis. \\ 
\noalign{\global\arrayrulewidth=0.4pt}\arrayrulecolor{black}
\hline
& generateNavigation & IAeroData<AeroNav> & IAeroData<AeroNav> & Generates a sequence of waypoints based on the navigation scheduler logic. \\ 
\noalign{\global\arrayrulewidth=0.1pt}\arrayrulecolor{lightgray}
\cline{2-5}
INavigate & addNavigation & IAeroData<AeroNav> & -- & Adds new waypoints to the already existing queue. \\ \cline{2-5}
& clearNavigation & -- & -- & Clears the existing waypoints from the queue. \\ 
\noalign{\global\arrayrulewidth=0.4pt}\arrayrulecolor{black}
\hline

INavigableAnalyse & generateNavigation & IAeroData & IAeroData<AeroNav> & Generates navigation commands based on analytics. \\

\noalign{\global\arrayrulewidth=0.4pt}\arrayrulecolor{black}
\hline

\end{tabular}
\end{table*}

\subsection{\adp APIs}
\adp provides a set of API interfaces that enable developers to orchestrate analytics and navigation services, manage UAV fleets, and integrate custom scheduling or path-planning modules with ease. These are discussed in detail.

\subsubsection{Environment Management Interface (\texttt{IEnviron\-ment})}
This provides unified access to environmental resources such as sensors, drones, and compute devices. These can be initialized from a configuration file or dynamically discovered. Developers can query sensor and robot (drone) availability (e.g., \textit{hasEnvSensors()}, \textit{getEnvSensorByID()}, \textit{getRobotByID()}) present in the environment. Additionally, compute resources that are accessible within the environment can be accessed through \textit{getComputeResources()}, and specific ones can be fetched using \textit{getComputeResourceByID()}. These may be edge devices onboard the drone, nearby edge accelerators, or remote cloud resources. The \textit{setTrajectoryScheduler()} and \textit{setAnalyticsScheduler()} can be used to plug in custom waypoints schedulers and analytics schedulers in the drone ecosystem, respectively. If these APIs are not used, \adp uses the default schedulers.
This enables adaptive, responsive drone applications. While our current work implements core functionality, future efforts will allow initializing \textit{IEnvironment} using a drone service provider’s URL to access remote resources seamlessly.

\subsubsection{Mission Control Interface (\texttt{IRobot})}
This enables hardware-agnostic mission execution and control of drone operations. It supports lifecycle interfaces such as \textit{startMission()} and \textit{endMission()} for custom logic, as well as \textit{pauseMission()} and \textit{resumeMission()} to temporarily halt and resume missions. The availability of onboard sensors are first verified using \textit{hasSensors()}, and, if available, can be accessed through \textit{getSensors()} or retrieved individually using \textit{getSensorByID()} for a specific sensor type. Navigation-related services include \textit{isNavigable()}, which determines whether a robot supports navigation, and \textit{navigate(IAeroData<AeroNav>)} to execute navigation commands based on provided parameters. In future, this can even be extended to ground robots alongside UAVs.

\subsubsection{Sensor Data Access Interface (\texttt{ISensor})}
This enables seamless interaction with onboard and environmental sensors. Applications can query sensor attributes using \textit{getSensorProperty(prop)} and identify sensors via \textit{getSensorID()}. Real-time data acquisition is supported through \textit{getDataStream()}, returning continuous \textit{AeroStreamData}. It supports flexibility in data handling using \textit{IPushSensor} for callback-based subscriptions and \textit{IPullSensor} for periodic polling.

We also design a new sensor, \textit{stat-stream}, which implements the \textit{ISensor} interface. It is a software-based odometry sensor modeled as a finite state machine that tracks and reports the current mission state of the drone. This sensor provides reliable state feedback, enhancing mission robustness and improving tolerance to odometry inaccuracies.

\subsubsection{Data Management Interface (\texttt{IAeroData})}
This standardizes data handling in \adp through the generic \textit{AeroData<E>} type, enabling consistent manipulation of diverse data such as images, sensor readings, and analysis results. It provides \textit{getData()} and \textit{setData(E data)} methods for accessing and modifying encapsulated values, supporting structured, unstructured, and custom formats. Extensions like \textit{AeroStreamData}, used by \textit{ISensor}, provide continuous data streams, while \textit{AeroListData} enables managing collections, such as navigation waypoints in \textit{AeroNav}.

\subsubsection{Compute Management Interface (\texttt{ICompute} \&  \texttt{IDeployable})}
\textit{ICompute} manages compute resources across the edge-cloud continuum and enables initialization and access to analytics services. Developers can retrieve device capabilities using \textit{getComputeProperties()}, including details like processing power, memory, and accelerators. Analytics can be deployed via \textit{IDeployable}, which may internally use edge, cloud, or hybrid analytics schedulers initialised using \textit{setAnalyticsScheduler()} in \texttt{IEnvironment} to balance workloads. The design is extensible to support diverse compute and scheduling strategies.

\subsubsection{Navigation Management Interface (\texttt{INavigate})}
This design enables seamless integration of navigation functions for dynamically managing and updating the drone's trajectory. Developers can initialize the initial route by invoking the \textit{generateNavigation()} method, which takes a user-defined set of waypoints and applies the scheduling logic defined in \textit{setTrajectoryScheduler()} within \texttt{IEnvironment} to generate an optimized waypoint sequence. These waypoints are then inserted into a queue for execution. The \textit{navigate()} function in \texttt{IRobot} subsequently polls waypoints from this queue to control the robot’s movement. New waypoints can be dynamically inserted during the mission using \textit{addNavigation()}, which supports appending at custom indices, thereby enabling mid-mission updates and priority-based scheduling. Meanwhile, the \textit{clearNavigation()} method resets the queue by removing all existing waypoints, allowing the drone to transition to a new mission when required. Collectively, these methods provide a flexible and efficient mechanism for managing drone trajectories across diverse mission patterns.

\begin{figure}[t]
    \centering
    \includegraphics[width=1\columnwidth]{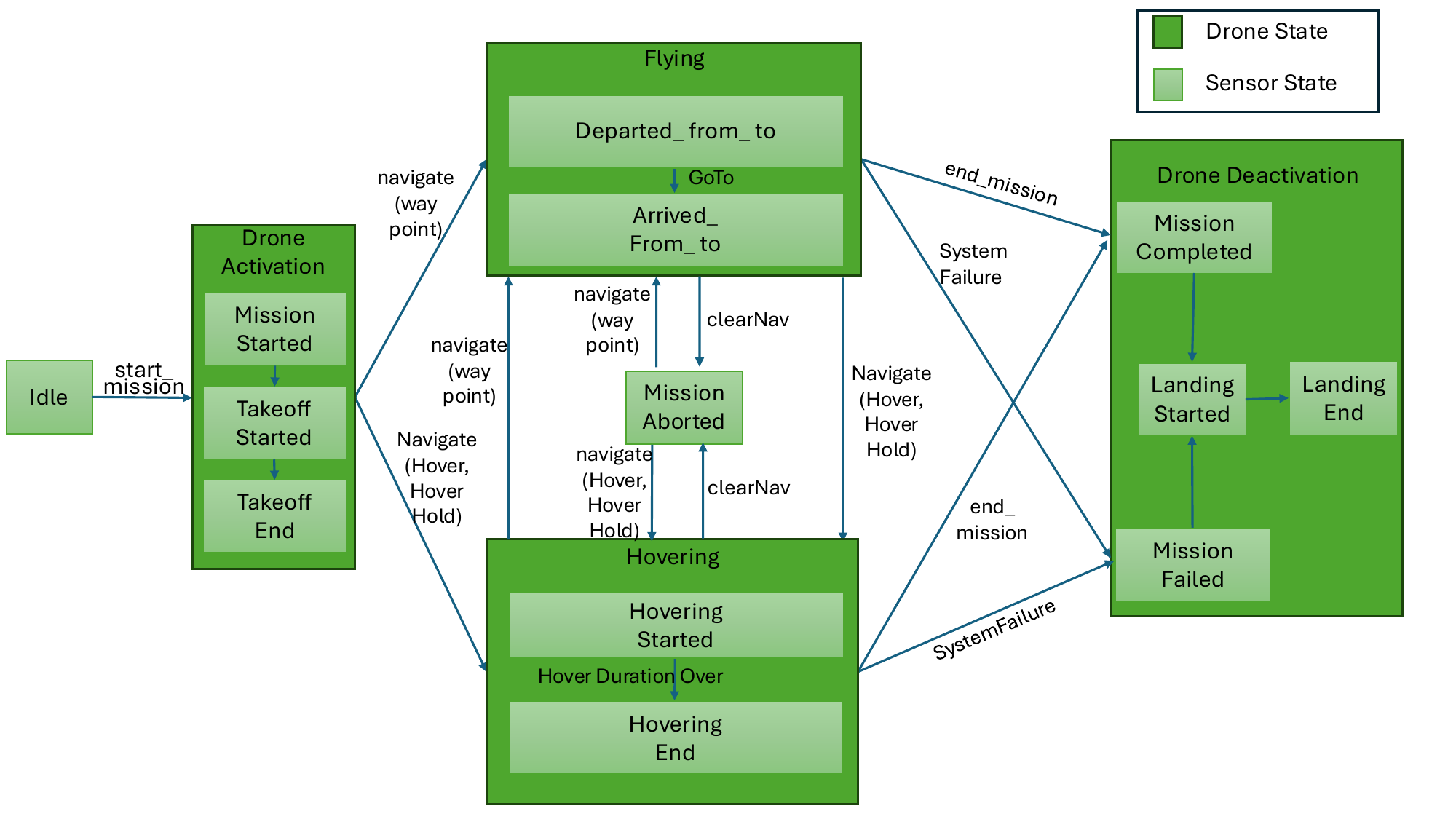}
    \caption{State transition diagram for Stat-Stream sensor}
    \label{fig:std-stat-stream}
\end{figure}

\subsubsection{Analytics and Analytics-based Navigation Interface (\texttt{IAnalyse} and \texttt{INavigable\-Analyse})}
\textit{IAnalyse} enables analytics-driven decision-making by allowing applications to invoke \textit{analyse(IAeroData)} for AI-based insights, such as performing object detection and returning bounding boxes over a video stream input. Data from external sensors can also be incorporated into the decision-making process using the \textit{instantiate()} method. Additionally, the \textit{trigger()} method enables the dynamic invocation of other analytics modules during mission execution. These triggers can originate from the output of existing analytics or be initiated dynamically by a remote user. Furthermore, \textit{INavigableAnalyse} extends these capabilities through \textit{generateNavigation(AeroData)}, which produces \textit{AeroData <AeroNav>} outputs, e.g., generating navigation directives from real-time sensor streams. Together, these interfaces facilitate seamless integration of DNN models for real-time analysis and control, thereby enhancing the system’s autonomy.
Users can also implement custom analytics modules, such as \textit{IMonitoringAnalytics}, which implements \textit{IAnalyse} and supports visualization of parameters like battery status and odometry data. When integrated with the \textit{stat-stream} sensor, it serves as a feedback mechanism to reconcile the drone’s perceived navigation progress with its actual odometry readings. Fig.~\ref{fig:std-stat-stream} illustrates an example of the drone’s mission states, where the \textit{stat-stream} sensor defines $13$ distinct states (depicted in light-green boxes) that collectively represent the full mission cycle across various possible mission patterns.

\section{Workflows}
\label{sec:workflows}
\begin{figure}
    \centering
    \includegraphics[width=0.75\linewidth]{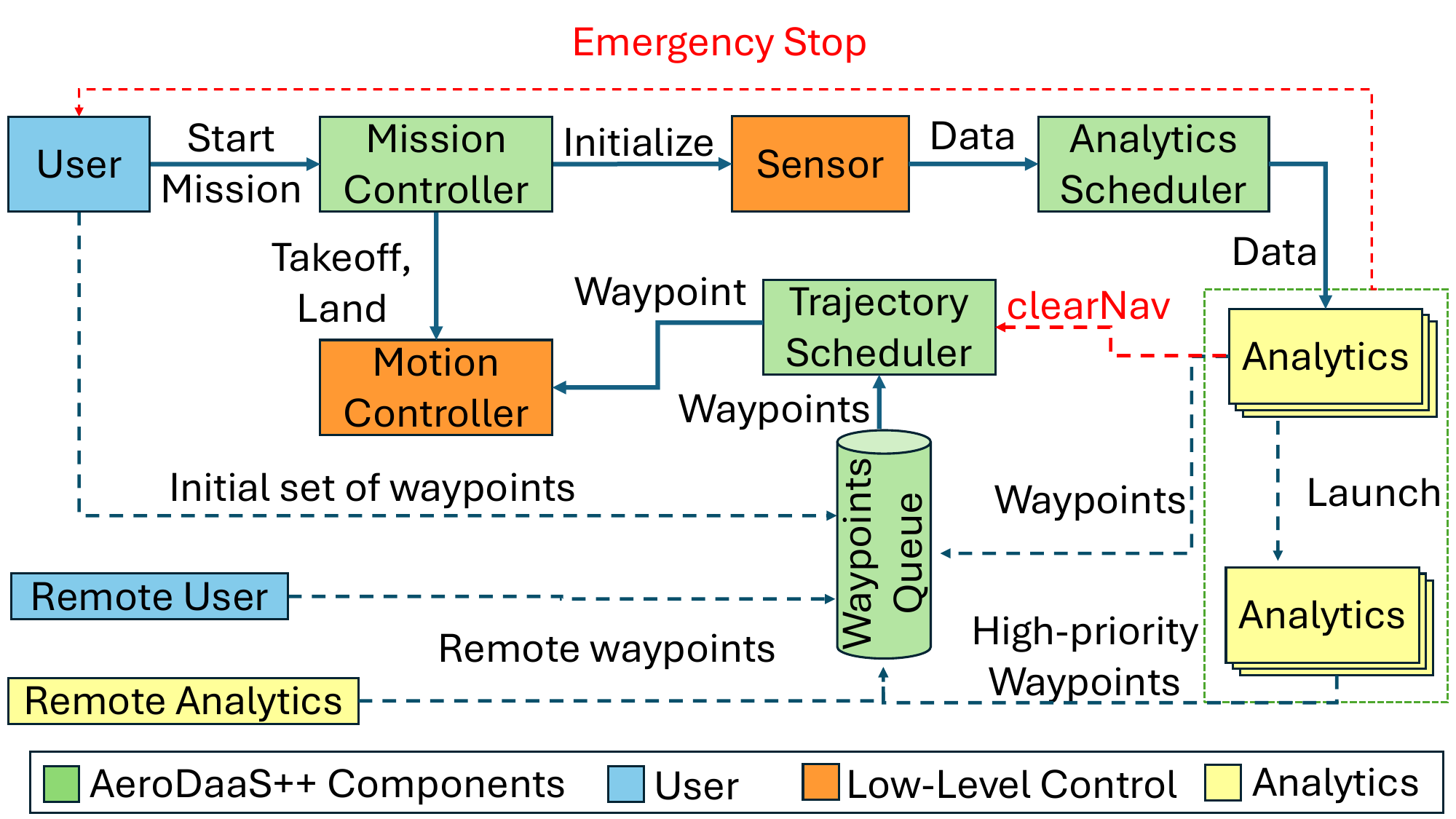}
    \caption{\adp workflow for different mission patterns}
    \label{fig:workflow-combined}
\end{figure}

In this section, we discuss various mission workflows in detail based on the mission patterns described in Sec.~\ref{subsec:mission-pattern}. Across all workflows, the mission is initiated by the user, following which the mission controller issues appropriate commands to the motion controller for drone takeoff and subsequently initializes the onboard sensors. Once activated, the sensors begin streaming data to the analytics scheduler, which coordinates the execution of onboard analytics modules. A waypoint priority queue collects waypoints originating from multiple sources with varying levels of priority. These waypoints are processed by the trajectory scheduler, which selects and dispatches the next waypoint to the motion controller, guiding the drone’s movement to the next location or initiating landing operations as required. An overview of all the workflows is illustrated in Fig.~\ref{fig:workflow-combined}, where each mission pattern has a distinct source for waypoint generation.

\subsection{Pre-defined Mission (PDM) with Static Waypoints}
\label{sec:worfklow-static}
In this workflow, the user initializes the mission using a predefined set of static waypoints. These waypoints are processed by the trajectory scheduler and sequentially dispatched to the motion controller for execution. Meanwhile, onboard sensors may stream data to the analytics module for task-specific processing. This workflow supports deterministic operations suited for pre-planned missions that require repeatability and precision, such as surveying or area mapping, without any real-time trajectory modification.

\subsection{Sensor Data-driven Mission}
\label{sec:workflow-sensor}
In this workflow, the mission begins without any predefined waypoints. Instead, the analytics module dynamically generates waypoints in real time based on incoming sensor (e.g., video) data and continuously inserts them into the waypoint queue. The trajectory scheduler processes these waypoints and forwards them to the motion controller for execution. As a result, the entire navigation process becomes fully autonomous and adaptive, guided by sensor-driven analytics.

\subsection{Pre-defined Mission (PDM) with Dynamic Waypoints}
This workflow starts with a predefined set of waypoints, similar to \S~\ref{sec:worfklow-static}, but allows dynamic modifications during execution. New waypoints can be inserted into the queue, either by remote analytics or user intervention, based on live sensor data or environmental feedback. The trajectory scheduler reprocesses the updated waypoints and sends them to the motion controller in sequence, enabling responsive and flexible mission adaptation in real time.

\subsection{Analytics-driven mission update}
\label{sec:workflow-mission-update}
In this workflow, the mission begins similarly to \S~\ref{sec:workflow-sensor}. However, based on onboard analytics output, a new analytics module can be dynamically launched, which generates high-priority waypoints based on detected events or real-time analytical insights (e.g., anomaly detection or target identification). These analytics-generated waypoints are inserted into the existing waypoint queue with elevated priority, enabling the drone to temporarily deviate from its ongoing mission to address critical tasks. Once the high-priority task is completed, the trajectory scheduler automatically restores the previous mission plan, allowing the drone to resume its original route seamlessly. This workflow exemplifies closed-loop autonomy where the \textit{Analytics} layer directly influences navigation decisions through data-driven triggers, ensuring responsiveness to dynamic and unpredictable environments.

\subsection{Analytics-driven mission abort}
In this workflow, onboard analytics are granted full authority to override the ongoing mission upon detecting or triggering a critical event, such as an obstacle, hazard, or emergency condition. When such an event occurs, the analytics module issues a \textit{clearNavigation} signal to the trajectory scheduler, immediately clearing all existing waypoints from the navigation queue. Unlike \S~\ref{sec:workflow-mission-update}, the drone does not resume its prior mission once the abort signal is received. This workflow demonstrates a robust fail-safe mechanism that prioritizes mission safety and real-time adaptability, ensuring the system can autonomously respond to unforeseen circumstances.

\section{Developing Applications using \adp}
\label{sec:application-psuedocodes}

\begin{figure*}[t]
    \centering
    \includegraphics[width=1.0\columnwidth]{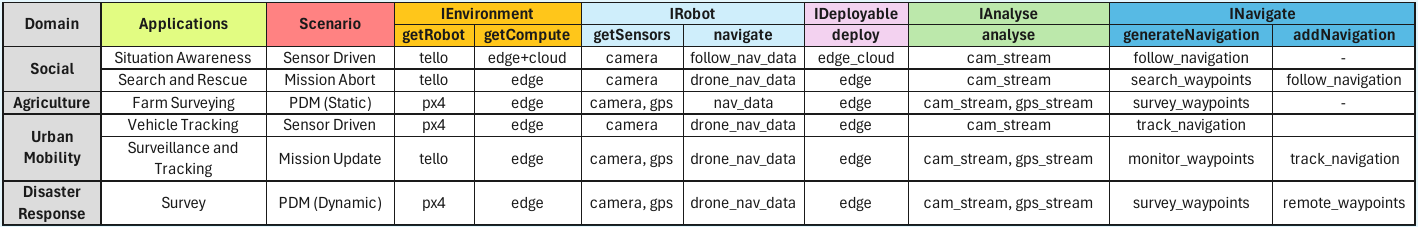}
    \caption{Diverse applications composed using various interfaces offered by \adp}
    \label{fig:applications-primitives-table}
\end{figure*}

In this section, we use the \adp APIs to compose the DaaS applications from our motivating use cases. We provide representative code snippets that use the programming and data interfaces to compose application services as dataflows. Fig.~\ref{fig:applications-primitives-table} illustrates the coverage of our APIs by the applications.
The environment initialization bootstraps the resources available to the developer. For simplicity, we assume obstacle avoidance is enabled natively by default for the drones.

\subsection{Situation Awareness}
\label{sec:application-psuedocodes:vip}
We show an analytics-driven (specifically, sensor-driven) application where drones are used to assist visually impaired people (VIP) for navigation~\cite{raj2023ocularone}. Here, the drone's camera stream is processed by a Hazard Vest Detection module based on YOLOv8-nano. The module generates bounding boxes around the vest, which are then translated into navigation commands using a PID controller, allowing the drone to autonomously track the VIP. Simultaneously, the same camera feed is analyzed by a Body Pose Analytics module that leverages a ResNet18-based DNN in combination with an SVM classifier to identify any gestures requiring responsive actions, such as commanding the drone to land when the VIP raises a hand or detecting a fall event for immediate assistance.

\begin{lstlisting}[language=python]
env: IEnvironment = Environment("environment.json")
edge_compute: ICompute = env.getComputeResourceById("edge_cloud")
drone: IRobot = env.getRobotById("tello")(*@\label{lst:vip-setup-camera-start}@*)
scheduler = NearestNeighborScheduler(env)
env.setTrajectoryScheduler(scheduler)
camera: IPushSensor[Image] = drone.getSensorById("tello_cam1")
camera_data_stream: AeroStreamData[Image] = camera.getDataStream()(*@\label{lst:vip-setup-camera-end}@*)
# Detect VIP from camera feed
vip_detect_analytics: IAnalyse = VipDetectionAnalytics()(*@\label{lst:vip-detect-start}@*)
vip_detect_analytics.deploy(edge_compute)
vip_detect_data_stream: AeroStreamData[Bbox] = vip_detect_analytics.analyse(camera_data_stream)
# Follow VIP based on bounding box
vip_follow_analytics: INavigableAnalyse = VipFollowAnalytics()
vip_follow_analytics.deploy(edge_compute)
vip_nav_data: IAeroData[AeroNavigation] = vip_follow_analytics.analyse(vip_detect_data_stream)(*@\label{lst:vip-detect-end}@*)
# Identify bodypose of VIP to send alert
vip_body_pose_analytics: IAnalyse = BodyPoseEmergencyAnalytics()
vip_body_pose_analytics.deploy(edge_compute)
vip_body_pose_data_stream: AeroStreamData[Bbox] = vip_body_pose_analytics.analyse(camera_data_stream)
# Send navigation commands to drone
drone_nav_data: AeroNavigation = scheduler.generateNavigation(vip_nav_data)
drone.navigate(drone_nav_data)
drone.start_mission()         
\end{lstlisting}

\subsection{Farm Survey}
\label{sec:application-psuedocodes:farm-survey-scene1}
Here, we present a waypoint-driven application in which a PX-4 controller drone called \texttt{quadrotor\_base} is initialized to survey a predefined rectangular or polygonal area, with boundary waypoints specified by the user. During the mission, the drone’s GPS data and camera video feed are continuously streamed to downstream applications, which can either store the data locally or upload it to the cloud for further processing and analysis.

\begin{lstlisting}[language=python]
env: IEnvironment = Environment("environment.json")
edge_compute: ICompute = env.getComputeResourceById("edge")
drone: IRobot = env.getRobotById("quadrotor_base")
scheduler = NearestNeighborScheduler(env)
env.setTrajectoryScheduler(scheduler)
odometry: IPushSensor = drone.getSensorById("odom")
odometry_data_stream: AeroStreamData[Odom] = odometry.getDataStream()
# Drone odometry monitoring
odometry_analytics: IAnalyse = MonitoringAnalytics(["battery", "height", "gps"])
odometry_analytics.deploy(edge_compute)
_: AeroStreamData[Odom] = odometry_analytics.analyse(odometry_data_stream)
wp1: Waypoint = Waypoint(id="wp1", x=20, y=20, z=10, hover_duration=5, waypoint_type="relative")(*@\label{lst:farm-wp1}@*)
wp2: Waypoint = Waypoint(id="wp2", x=20, y=100, z=10, hover_duration=5, waypoint_type="relative")
wp3: Waypoint = Waypoint(id="wp3", x=60, y=100, z=10, hover_duration=5, waypoint_type="relative")
wp4: Waypoint = Waypoint(id="wp4", x=60, y=20, z=10, hover_duration=5, waypoint_type="relative")
waypoints_list: AeroNavigation = AeroNavigation(NavigationType.DISTANCE_DRIVEN, [wp1, wp2, wp3, wp4], SchedulingType.UNORDERED)
drone_nav_data: IAeroData[AeroNavigation] = scheduler.generateNavigation(waypoints_list)(*@\label{lst:farm-wp-list}@*)
drone.navigate(drone_nav_data)
drone.start_mission()
\end{lstlisting}

\subsection{Human-in-the-loop Disaster Zone Survey}
\label{sec:application-psuedocodes:disaster-survey-scene2}
Here, we demonstrate a waypoint-driven Human-in-the-Loop application in which both a remote user and an onboard analytics module collaboratively control the drone. The mission begins with a predefined survey area specified through boundary waypoints provided by the user. Throughout the mission, the drone’s GPS telemetry and camera video feed are continuously streamed to downstream analytics modules, enabling real-time monitoring, adaptive decision-making, and interactive user intervention when required.

\begin{lstlisting}[language=python]
# Initialize env, edge resource, PX4-based drone
camera: IPushSensor = drone.getSensorById("cam1")
cam_data_stream: AeroStreamData[Image] = camera.getDataStream()
odometry: IPushSensor = drone.getSensorById("odom")
odometry_data_stream: AeroStreamData[Odom] = odometry.getDataStream()
# Drone odometry monitoring
odometry_analytics: IAnalyse = MonitoringAnalytics(["battery", "height", "gps", "camera"])
odometry_analytics.deploy(edge_compute)
_: AeroStreamData[Odom] = odometry_analytics.analyse(odometry_data_stream)
Static Waypoint list generation as (*@\ref{lst:farm-wp1}@*) - (*@\ref{lst:farm-wp-list}@*) from (*@\ref{sec:application-psuedocodes:farm-survey-scene1}@*)
remote_analytics: IAnalyse = RemoteAnalytics(env.env_data)
remote_analytics.deploy(edge_compute)
remote_nav_data: AeroStreamData = remote_analytics.analyse(cam_data_stream)
remote_nav_data: AeroNavigation = AeroNavigation(NavigationType.ANALYTICS_DRIVEN, remote_nav_data.data, SchedulingType.ORDERED)
scheduler.addNavigation(remote_nav_data)
drone.navigate(drone_nav_data)
drone.start_mission()
\end{lstlisting}

\subsection{Surveillance and Tracking}
\label{sec:application-psuedocodes:vehicle-tracking-scene4}
In this application, we present a hybrid waypoint and analytics-driven application, where the mission begins with a predefined area-surveillance task using boundary waypoints. During flight, the drone’s camera stream is continuously analyzed, and upon detecting a specific target vehicle, the system transitions into a tracking mode. The video feed is processed by the \textit{CarFollow} analytics, which uses a YOLOv11-medium model to identify the target and generate navigation commands for vehicle tracking. Once the vehicle exits the surveillance region, the drone autonomously reverts to its survey mission.

\begin{lstlisting}[language=python]
# Initialize env, edge resource, tello drone
Get Tello camera stream like (*@\ref{lst:vip-setup-camera-start}@*)-(*@\ref{lst:vip-setup-camera-end}@*) from (*@\ref{sec:application-psuedocodes:vip}@*)
wp1 : Waypoint = Waypoint(id = "wp1", waypoint_type="relative", x=20, y=20, z=10)
wp2 : Waypoint = Waypoint(id = "wp2", waypoint_type="relative", x=40, y=20, z=10)
wp3 : Waypoint = Waypoint(id = "wp3", waypoint_type="relative", x=60, y=20, z=10)
waypoints_list: AeroListData = AeroListData([wp1, wp2, wp3])
low_priority_queue: PriorityQueue[AeroListData] = PriorityQueue(waypoints_list, priority=2)
drone_nav_data: IAeroData[AeroNavigation] = scheduler.generateNavigation(low_priority_queue)
car_analytics: IAnalyse = CarFollowAnalytics(env)
car_analytics.deploy(edge_compute)
car_nav_data: IAeroData[AeroNavigation] = car_analytics.analyse(camera_data_stream)
car_nav_queue: PriorityQueue[AeroNavigation] = PriorityQueue(car_nav_data, priority=1)
scheduler.addNavigation(car_nav_queue)
drone.navigate(drone_nav_data)
drone.start_mission()
\end{lstlisting}

\subsection{Search and Rescue}
\label{sec:application-psuedocodes:search-rescue-scene5}
In this application, we demonstrate a Search and Rescue mission that integrates both waypoint-based and analytics-driven control within a Mission Abort scenario. The drone initially performs an autonomous search over a predefined area defined by user-provided boundary waypoints. During the operation, if a person wearing a hazard vest is detected, the onboard analytics triggers a mission abort, causing the drone to immediately terminate its current search and transition into a tracking mode. The live camera feed is then routed to the \textit{VipFollowAnalytics} module, which uses a YOLOv8-nano model to detect the hazard vest and generate real-time navigation commands, enabling the drone to closely follow the detected individual.

\begin{lstlisting}[language=python]
# Initialize env, edge resource, tello drone
Get Tello camera stream like (*@\ref{lst:vip-setup-camera-start}@*)-(*@\ref{lst:vip-setup-camera-end}@*) from (*@\ref{sec:application-psuedocodes:vip}@*)
Waypoint list generation same as (*@\ref{lst:farm-wp1}@*) - (*@\ref{lst:farm-wp-list}@*) from (*@\ref{sec:application-psuedocodes:farm-survey-scene1}@*)
Detect vip in the area like (*@\ref{lst:vip-detect-start}@*)-(*@\ref{lst:vip-detect-end}@*) from (*@\ref{sec:application-psuedocodes:vip}@*)
# On detecting vip, reset the queue and start vip following mission
scheduler.clearNavigation() # Reset the queue
scheduler.addNavigation(vip_nav_data)
drone.navigate(drone_nav_data)
drone.start_mission()
\end{lstlisting}

\section{Runtime Implementation}
\label{sec:runtime}

\begin{figure}[t]
    \centering
    \includegraphics[width=0.6\columnwidth]{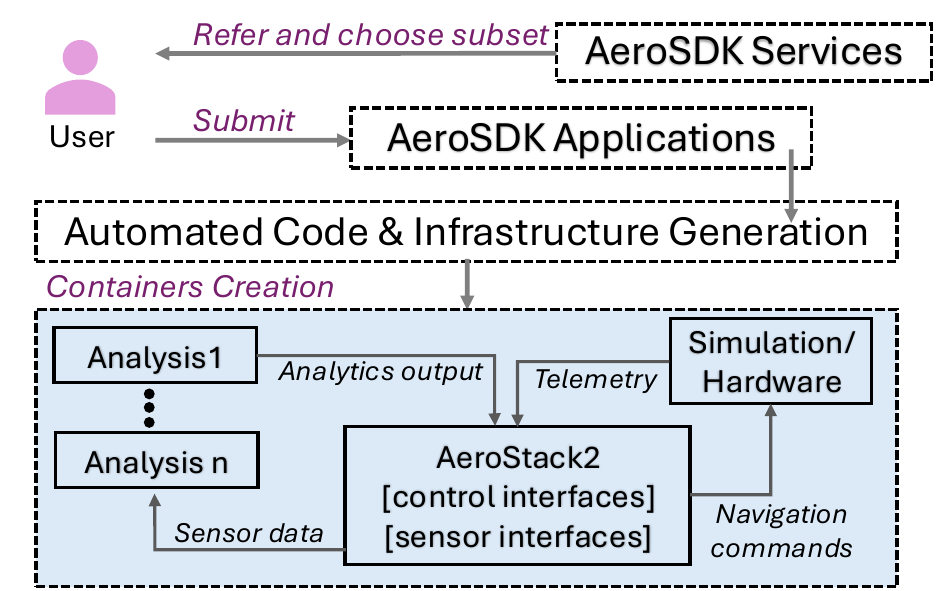}
    \caption{
    \adp Runtime Implementation}
    \label{fig:runtime-implementation}
\end{figure}

Fig.~\ref{fig:runtime-implementation} illustrates the key components of the \adp runtime implementation. The user interacts with the \adp APIs, selects a subset of available functions to define their application, and submits their application through a user-facing file. The following subsections detail the involved steps.

\subsection{Automated Code and Infrastructure Generation}
We implement our runtime in \textit{Python3} and provide equivalent Python APIs for the \adp services. Users develop applications as Python scripts, which execute within an isolated container. This container is set up using a pre-configured Docker image that includes (a) the AeroStack2 library, (b) backend support for the target drone hardware, and (c) the user’s transpiled application code embedded in a ROS2 node with sensor subscriptions and analytics interactions. The Docker image comes with all necessary dependencies pre-installed. 

\adp relies on Aerostack2 plugins as the hardware abstraction layer, requiring them to be active before any interface is used. When a user executes their Python application, it first invokes \textit{generate\_code}, the core function responsible for infrastructure setup. This function generates a Python-based ROS2 node file containing services for analytics, navigation, sensing, and compute, along with corresponding bash scripts to launch necessary services. The main file is dynamically populated based on user inputs using Jinja2~\cite{jinja2}. The bash scripts then initialize the required containers, including Aerostack2 services, analytics modules, and optional simulation environments if no physical drone is used. Finally, the generated ROS2 node executes, starting the mission.

\subsection{Containers Coordination using \adp Services}
AeroStack2 provides \textit{control interfaces} for drone navigation and \textit{sensor interfaces} for accessing onboard data, both exposed via ROS topics and messages. \adp ensures seamless inter-container communication by running all containers on the \textit{host} network, enabling data exchange over \textit{localhost}. \textit{ISensor} allows analysis containers to retrieve sensor data, while \textit{IAnalyse} facilitates transfer of analytics outputs from these containers to the AeroStack2 container. The \textit{navigate} function in \textit{IRobot} is responsible for transmitting navigation waypoints generated by \textit{INavigate} from the AeroStack2 container to either the simulation environment or the physical hardware robot. Similarly, telemetry data from the robot is received by the AeroStack2 container through \textit{IMonitoringAnalytics}, enabling real-time monitoring of system states and mission progress. The inter-container coordination mechanism remains identical in both real-world and simulated environments.

\section{Evaluation}
\label{sec:evals}

\begin{figure}
    \centering
    \subfloat[Nvidia Jetson Orin Nano]{
    \includegraphics[width=0.33\columnwidth,height=2.6cm]{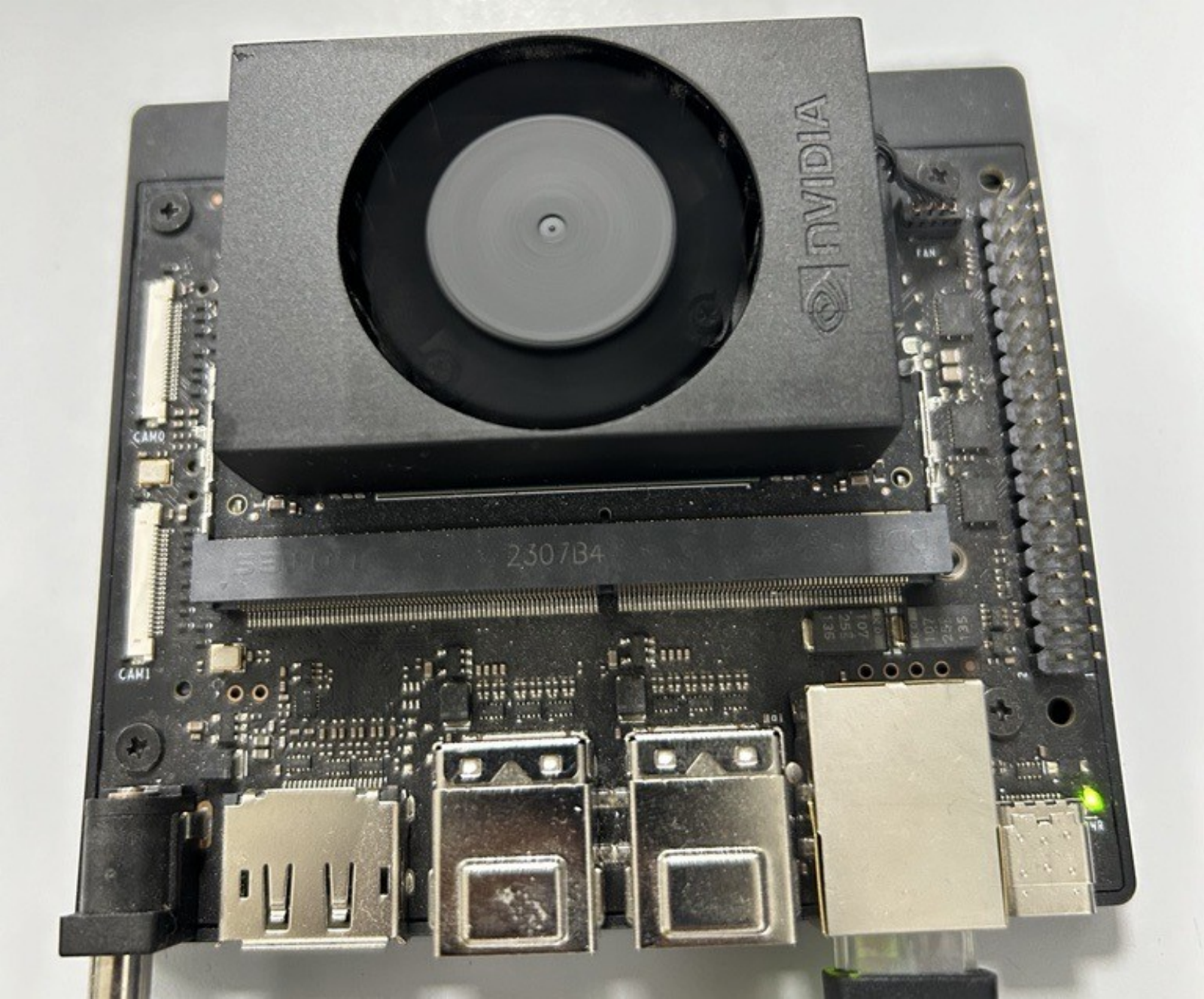}
    \label{fig:onano-hardware}
    }~~
    \subfloat[DJI Ryze Tello Drone]{
    \includegraphics[width=0.33\columnwidth,height=2.6cm]{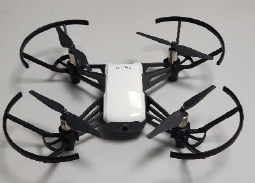}
    \label{fig:dji-tello-hardware}
    }~~
    \subfloat[Proxy VIP followed by Tello drone]{
    \includegraphics[width=0.25\columnwidth, height=2.6cm]{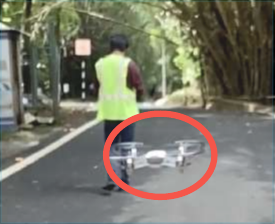}
    \label{fig:dummy-vip}
    }\\
    \subfloat[Quadrotor Base Drone]{
    \includegraphics[width=0.33\columnwidth, height=2.6cm]{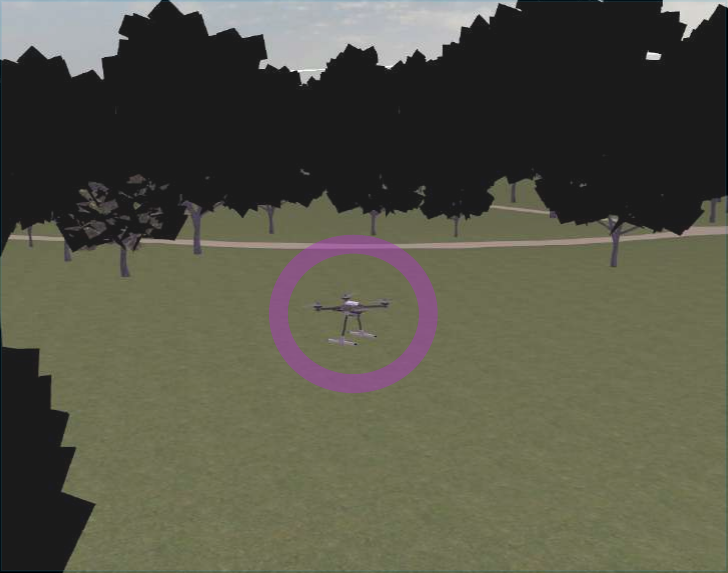}
    \label{fig:farm-survey-drone}
    }
    \subfloat[World with a dummy cell tower]{
    \includegraphics[width=0.28\columnwidth,height=2.6cm]{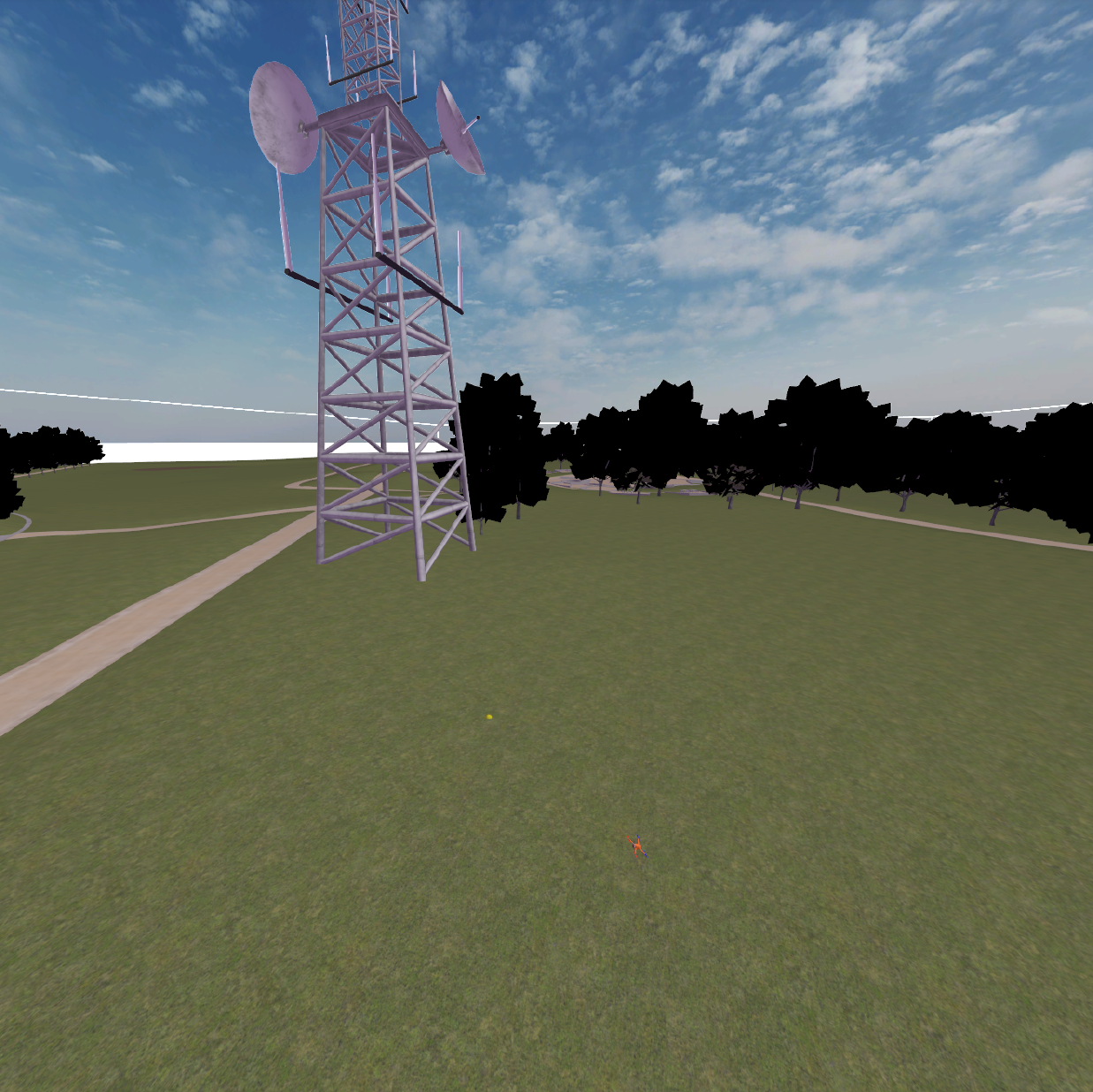}
    \label{fig:baylands_tower_world}
    }~~
    \subfloat[Prius Hybrid car model and drone (highlighted in blue circle)]{
    \includegraphics[width=0.28\columnwidth,height=2.6cm]{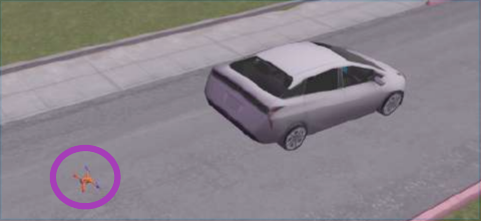}
    \label{fig:baylands_car_world}
    }
\caption{Experimental Setup}
\label{fig:exp-setup}
\end{figure}

We perform a detailed evaluation of \adp for five motivating applications: Farm Survey, Human-in-the-loop Disaster Survey, Situation Awareness, Search and Rescue, Surveillance and Vehicle Tracking.
These are implemented using our programming API and deployed within our runtime. These applications validate both real-world deployment with the Tello Drone using hardware-in-the-loop (HITL), and a simulation-based evaluation using a Quadrotor Base model drone supporting PX-4 controller featuring onboard GPS camera using software-in-the-loop (SITL). These demonstrate the flexibility of the framework to execute the application on different target platforms with optimized trajectory by integrating with a trajectory scheduler. We also evaluate the use of edge and cloud resources, individually and together, using a modular scheduler to execute analytics that drive the DaaS application.

\subsection{Setup}
For the HITL experiment, we use a DJI Ryze Tello drone and an Nvidia Jetson Orin Nano edge accelerator. The Orin Nano has a six-core Arm Cortex-A78AE CPU, 1024 Ampere CUDA cores, and 8GB of RAM shared by CPU and GPU (Fig.~\ref{fig:onano-hardware}). It has a power range of $7$-$15$W, is powered by a portable power bank with the max output of 12V/18W, has compact dimensions of $100 \times 79$mm, and connects to the drone over TP-Link TL-WN722N 150 Mbps high gain wireless USB WiFi adapter. The DJI Tello is equipped with an onboard camera capable of generating live video feeds at 30 FPS at $720p$ resolution (Fig.~\ref{fig:dji-tello-hardware}). For the situation awareness and search-and-rescue experiments, a proxy VIP (Fig.~\ref{fig:dummy-vip}) carries a smartphone with Android OS version 14. We use the GPS Logger Lite Android application~\footnote{\url{https://play.google.com/store/apps/details?id=com.peterhohsy.gpsloggerlite}} available on the Google Play Store, to record the VIP's current location during the experiments. The drone was maintaining a constant distance of 2m from the VIP throughout the experiment, and the experiments run on our university campus.

AeroStack2 and the required DNN models were executed inside Docker containers with base image of Ubuntu 20.04, Python 3.8.10 and PyTorch v2.0.0 and run on the NVIDIA Jetson Orin Nano for edge-based experiments, while Amazon Web Services (AWS) Lambda is utilized for cloud-based experiments. We follow the cloud experiment setup outlined in~\cite{RAJ2025107874}.
For SITL simulations, we use a Quadrotor base model drone that generates live video feeds at $15$~FPS (Fig.~\ref{fig:farm-survey-drone}) and use the \textit{Baylands} virtual world with a dummy cell tower (Fig.~\ref{fig:baylands_tower_world})  which, although lacking communication functionality, provides the necessary infrastructure for evaluating the Radio Tower Inspection application 
and Prius Hybrid car model~\footnote{\url{https://github.com/osrf/car_demo/}}(Fig.~\ref{fig:baylands_car_world}) to support Vehicle Tracking application. The simulations are run in Gazebo simulator (GZ Garden) integrated with ROS2 (Humble) running inside a Docker container on a GPU workstation equipped with an AMD Ryzen 9 3900X 12-core CPU, 24GB RAM, and a Nvidia RTX 3090 GPU having 128 GB VRAM.

\begin{table*}[!t]
\centering
\caption{Summary of workloads and experimental configurations.}
\label{tab:workload-config}
\scriptsize
\setlength{\tabcolsep}{1pt}
\begin{tabular}{C{2cm}|C{2cm}|c|C{2.5cm}|c|C{2.5cm}|c|c}
\hline
\textbf{Workload} & \textbf{Analytical Models Used} & \textbf{FPS} & \textbf{Setup} & \textbf{Duration} & \textbf{Area/Distance Covered} & \textbf{Drone Altitude} & \textbf{Drone Speed} \\ \hline
\hline
Farm Survey & -- & 15 & Gazebo (Bayland world) & 2min & 10m $\times$ 10m & 1.5m & 0.5 m/s \\ \hline
Disaster Survey & -- & 15 & Gazebo (Bayland world) & 2.5min & 10m $\times$ 10m (preplanned) and 10m $\times$ 20m (remotely injected) & 1.5m & 1.5 m/s \\ \hline
Situation Awareness & YOLOv8-nano (object-detection), GPS Logger & 5 &  Outdoor field (IISc Bangalore campus) & 4min & 50m & 1.5m & 1.0 m/s \\ \hline
Vehicle Tracking & YOLOv8-nano (vehicle detection) & 15 & Gazebo (Bayland world with Prius Hybrid car) & 5min & 50 m & 2.5m - 5m & 0.5m/s--1.0 m/s \\ \hline
Search and Rescue & YOLOv8-nano (hazard vest detection) & 5 & Outdoor field (IISc Bangalore campus) & 6min & 50m & 1.5m & 1.0 m/s \\ \hline
Radio Tower Inspection & YOLOv8-nano & 15 &  Gazebo (Bayland world with dummy cell tower) & 6min & 130m & 2m--35m & 0.5m/s--1.5m/s \\ \hline
\end{tabular}
\end{table*}

\subsection{Workloads}

\begin{figure*}[!t]
\centering
    \subfloat[pre-planned static mission]{
        \includegraphics[width=0.33\textwidth]{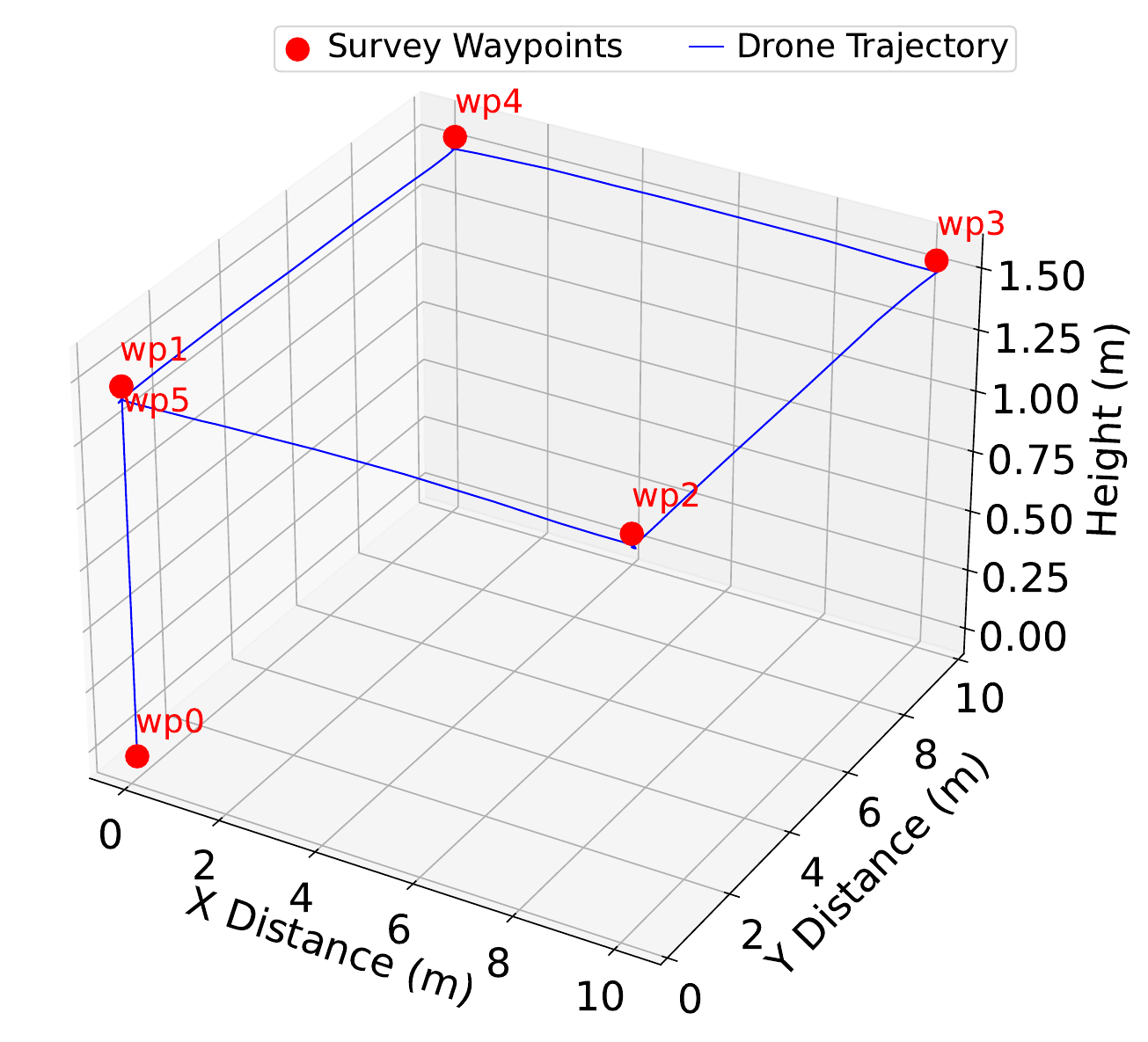}
        \label{fig:sim_scene1_trajectory}
    }
    \subfloat[mid-mission remote modification]{
        \includegraphics[width=0.33\textwidth]{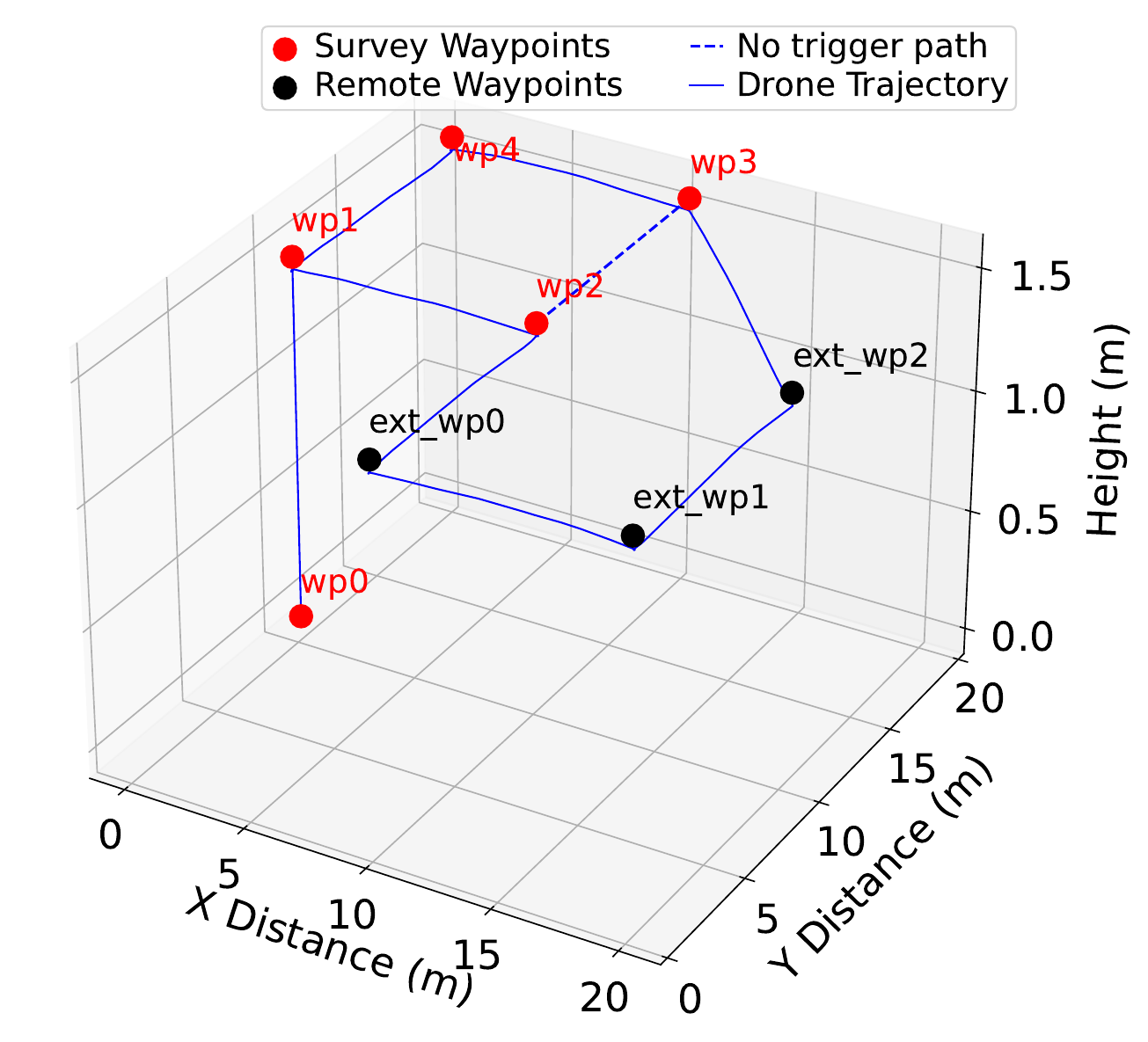}
        \label{fig:sim_scene2_trajectory}
    }
    \subfloat[sensor driven]{
        \includegraphics[width=0.33\textwidth]{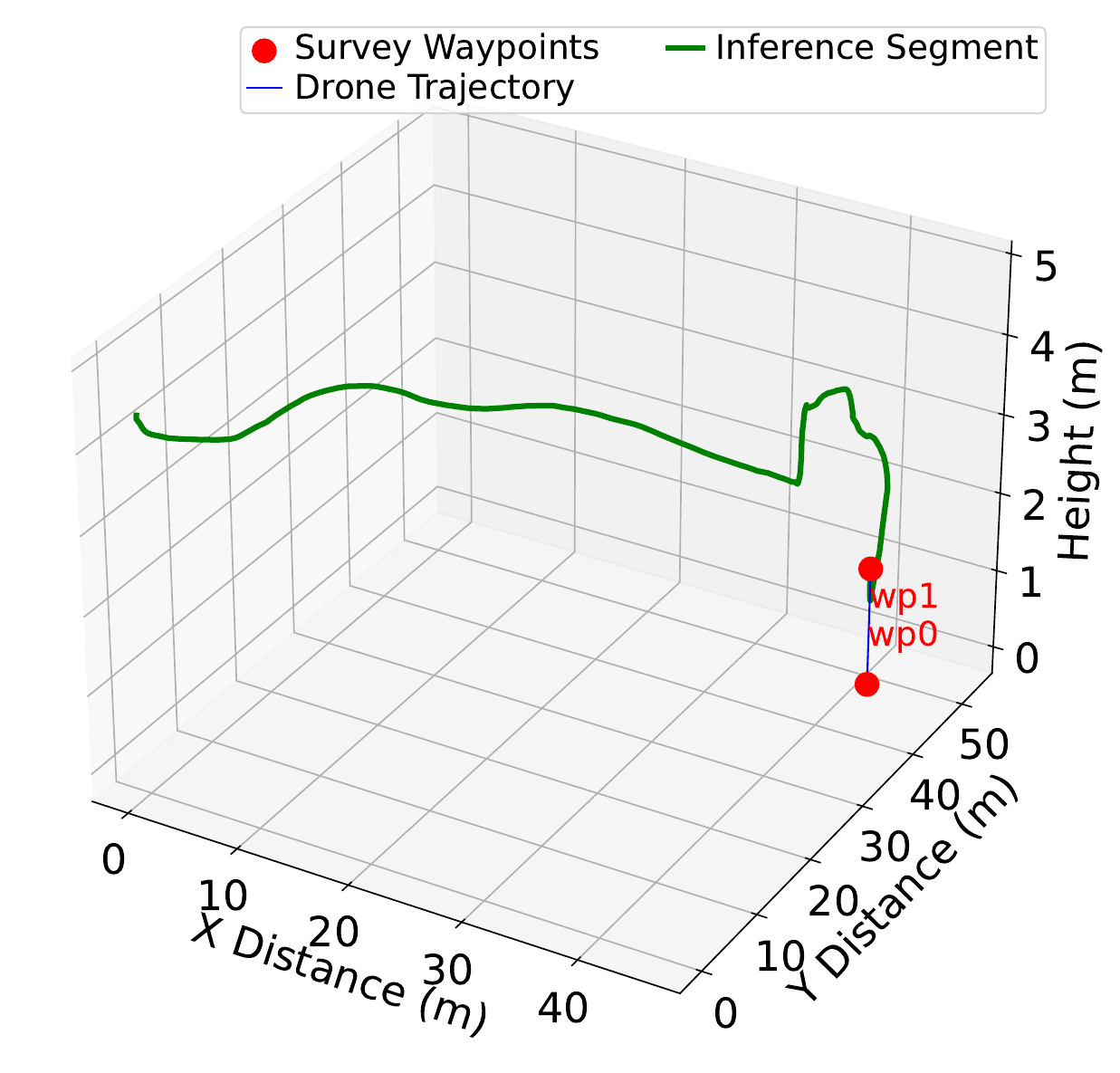}
        \label{fig:sim_scene3_trajectory}
    }\\
    \subfloat[mission change]{
        \includegraphics[width=0.33\textwidth]{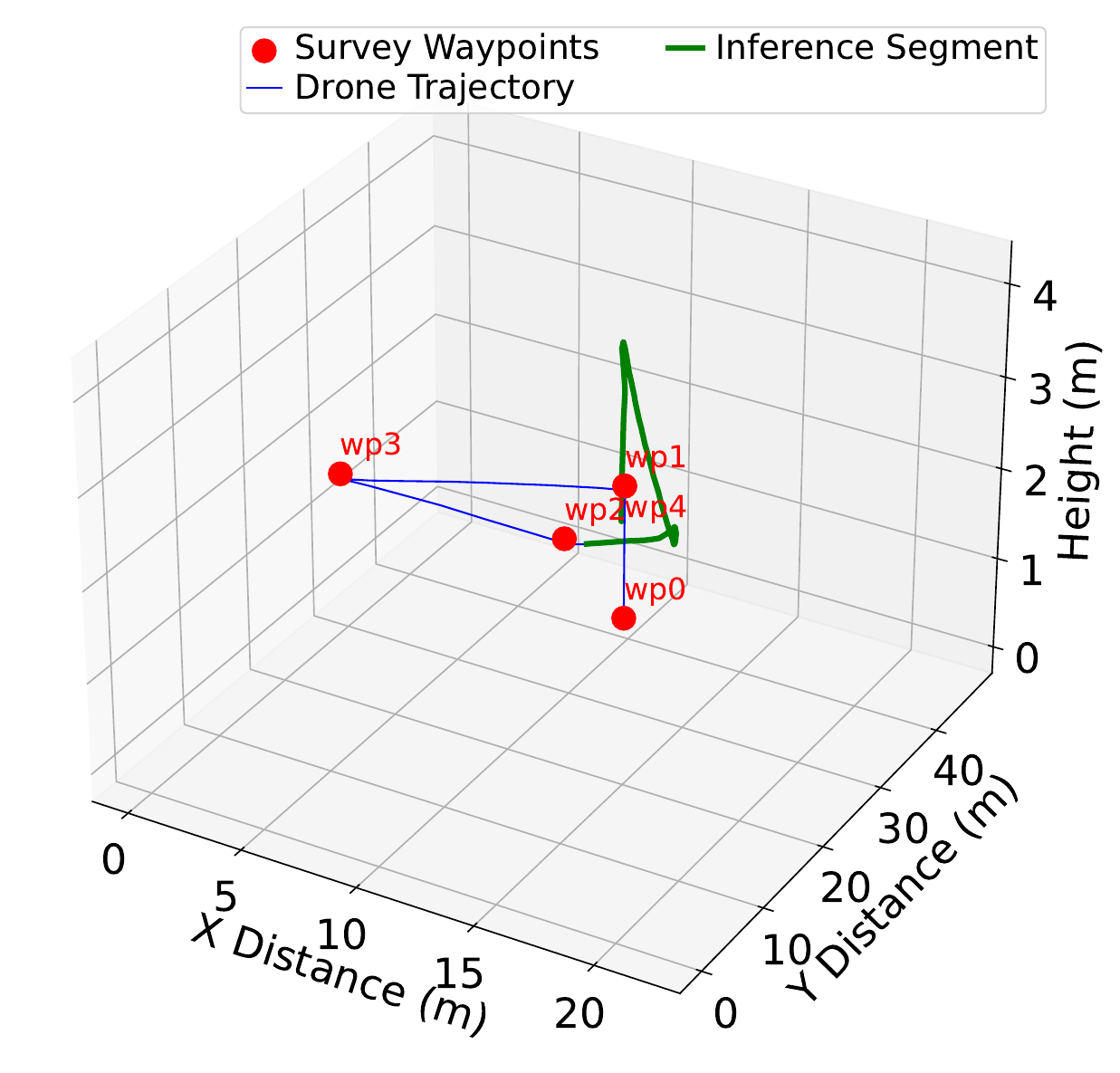}
        \label{fig:sim_scene4_trajectory}
    }
    \subfloat[mission abort]{
        \includegraphics[width=0.33\textwidth]{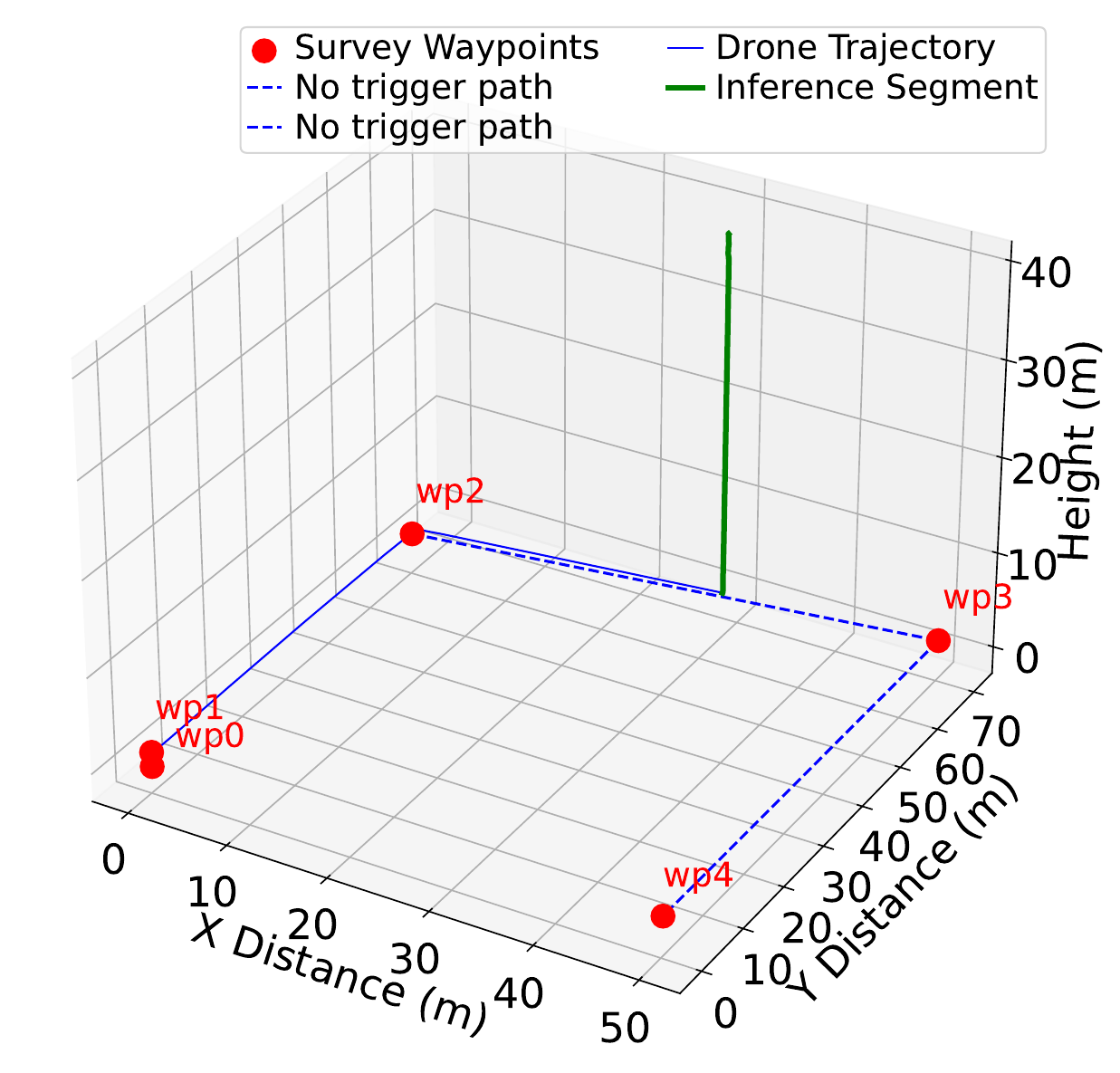}
        \label{fig:sim_scene5_trajectory}
    }
    \subfloat[Real-World Mission Abort]{
        \includegraphics[width=0.33\textwidth]{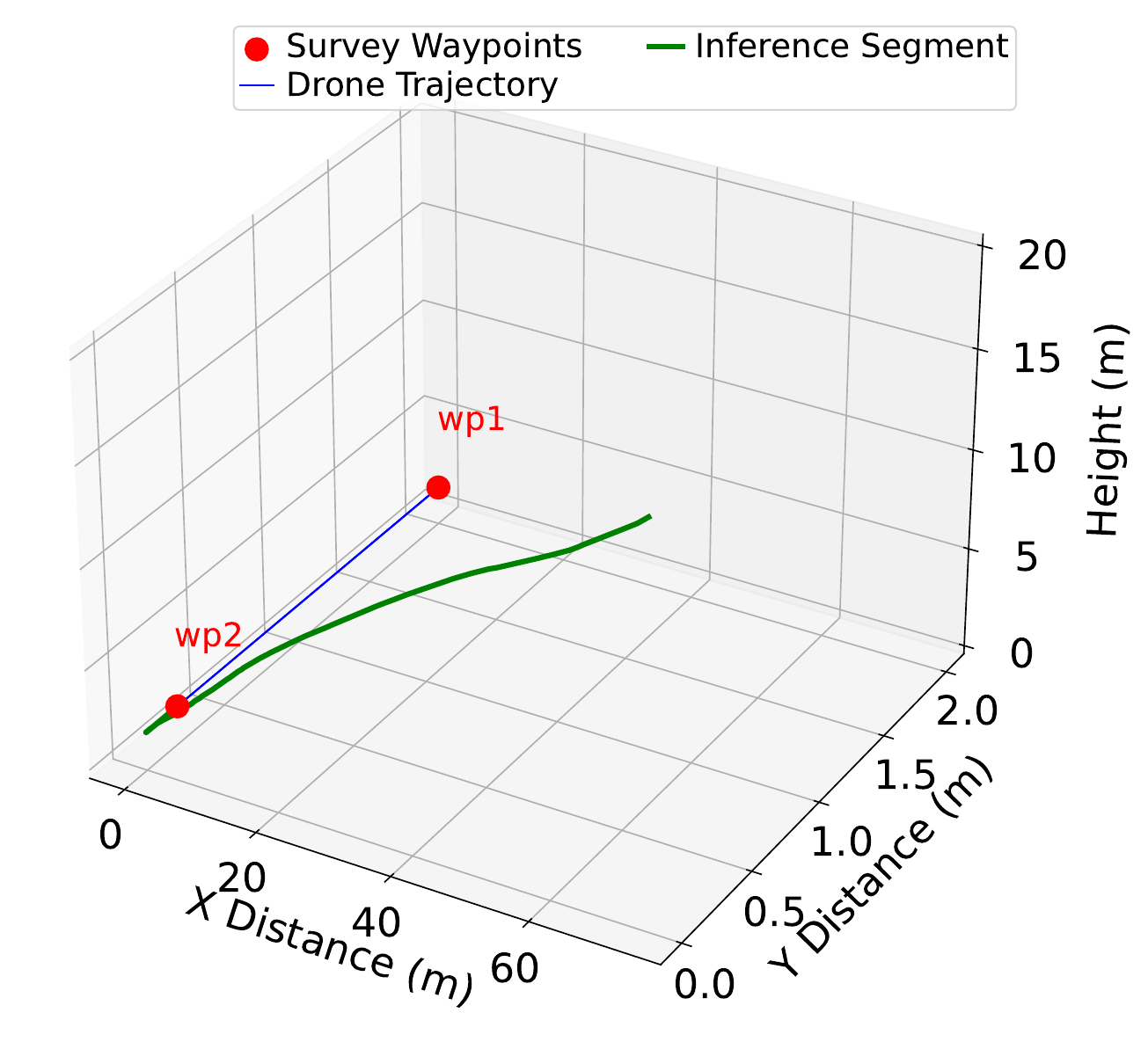}
        \label{fig:real_scene5_trajectory}
    }
\caption{Trajectory of simulated or real drone for different workloads}
\end{figure*}

We assess the performance of \adp using six representative applications that covers multiple UAV use case domains. A summary of these workloads is provided in Table~\ref{tab:workload-config} with the following detailed descriptions:

\subsubsection{Farm Survey}
\label{sec:workload-farm-survey}
This application demonstrates a pre-planned static mission designed to survey a small $10\text{m} \times 10,\text{m}$ area (Fig.~\ref{fig:sim_scene1_trajectory}). The experiment was conducted in Gazebo using the Baylands virtual world and a quadrotor base model where the drone was flown at an altitude of $1.5\text{m}$ with a speed of $0.5\text{m/s}$. The onboard camera streamed video at $15$ FPS, which was stored locally for offline analysis. At each waypoint, the drone hovered for $5$ seconds before proceeding to the next, completing the full survey in approximately $120$ seconds.

\subsubsection{Disaster Survey with Human Supervision}
\label{sec:disaster-survey}
This workload demonstrates a mid-mission modification pattern aimed at surveying a small disaster-affected region under human supervision. The mission begins with a pre-planned survey consisting of $4$ initial waypoints covering a $10\text{m} \times 10\text{m}$ area, followed by $3$ remotely injected waypoints covering a $10\text{m} \times 20\text{m}$ area (Fig.~\ref{fig:sim_scene2_trajectory}) to emulate dynamic task allocation during execution. The experiment was performed in simulation, where the drone was flying at an altitude of $1.5\text{m}$ for $142$ seconds at a speed of $1.5\text{m/s}$ while capturing video.

\subsubsection{Situation Awareness}
This workload demonstrates a sensor-driven mission aimed at assisting a VIP in an outdoor environment~\cite{raj2023ocularone}. In the HITL experiment, the drone was tasked with continuously following a VIP wearing a hazard vest.
It flew at an altitude of $1.5m$ with a variable speed ranging from $0.1m/s$ to $1m/s$ maintaining a constant distance of approximately $2m$ from the VIP. The onboard camera streamed video at $5$ FPS, which was processed using the fine-tuned YOLOv8-nano model for hazard vest detection. The detection results were post-processed to generate navigation speed commands for the drone. Simultaneously, the GPS Logger Lite application recorded the VIP's location data to support spatio-temporal analysis. The experiment lasted for approximately $250$ seconds.

\subsubsection{Surveillance and Vehicle Tracking}
This workload represents an analytics-driven mission update that integrates area surveillance with vehicle tracking. The experiment was conducted in Gazebo using the Baylands virtual world, with a Prius Hybrid car serving as the moving target and a quadrotor base model drone flying at an altitude of $2.5\text{m}$. The drone was initialized with a pre-planned survey mission consisting of three waypoints. During execution, it successfully completed the first waypoint before detecting a car through its onboard camera feed, processed using a fine-tuned YOLOv8-nano model. Upon detection, the drone dynamically transitioned from survey mode to vehicle-tracking mode and followed the car for approximately $30\text{m}$ (Fig.~\ref{fig:sim_scene4_trajectory}). The car moved at a low speed of $0.5\text{m/s}$ to enable reliable detection and tracking. Once the car exited the survey boundary taking a sharp right turn and moving out of the drone's FPV camera the drone autonomously resumed its original survey mission to complete the remaining waypoints. In the absence of any pre-planned waypoints, this workload naturally reduces to a pure vehicle tracking scenario following the sensor-driven paradigm. Fig.~\ref{fig:sim_scene3_trajectory} further illustrates a simulated drone tracking a car for more than $64\text{m}$ in Gazebo.

\subsubsection{Search and Rescue}
This workload represents an analytics-driven abort mission in real world where the drone is initialized with a search mission followed by a rescue mission. The experiment was performed in the real world with the drone initialized with $4$ waypoints to search for a VIP wearing a hazard vest. During the mission, the VIP was detected at the second waypoint using the onboard camera feed processed by the fine-tuned YOLOv8-nano model. Upon detection, the drone aborted its search mission and transitioned into the VIP following mode. The drone then followed the VIP for next $6$ minutes until the VIP reached a safe waypoint located $40m$ away from the initial position (Fig.~\ref{fig:real_scene5_trajectory}).

\subsubsection{Radio Tower Inspection}
This workload demonstrates an analytics-driven abort mission performed in simulation using the Gazebo simulator with the world having a dummy cell tower for inspection. The drone was initialized with $4$ waypoints to locate the cell tower. Upon identifying the tower between waypoints $2$ and $3$, the drone switched to an inspection mode, capturing images from the tower's base to an altitude of $35m$ (refer Fig.~\ref{fig:sim_scene5_trajectory}) and streaming them to the fine-tuned YOLOv8-nano model for analysis. After completing the inspection, the drone made a smooth landing.

\subsection{Results}
Next, we present key insights from our experiments that validate these six applications, implemented and orchestrated using \adp, each running for more than 250 seconds. 

\begin{figure}[!t]
\centering
\includegraphics[width=0.9\columnwidth]{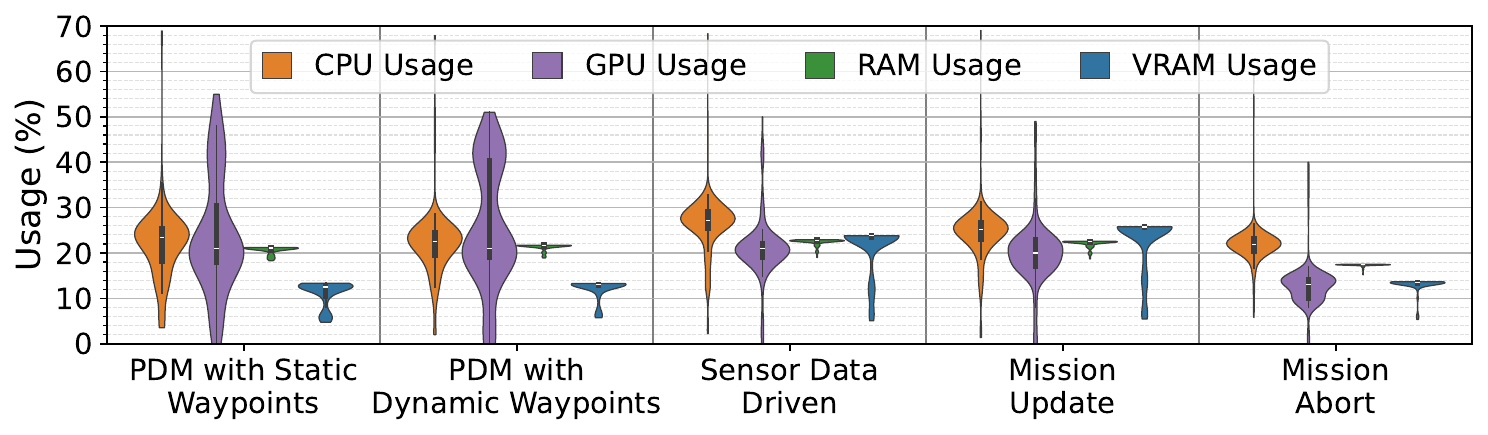}
\caption{System metrics during simulation of various mission pattern}
\label{fig:system-usage-sim}
\end{figure}

\begin{figure}[t]
\centering
\includegraphics[width=0.7\columnwidth]{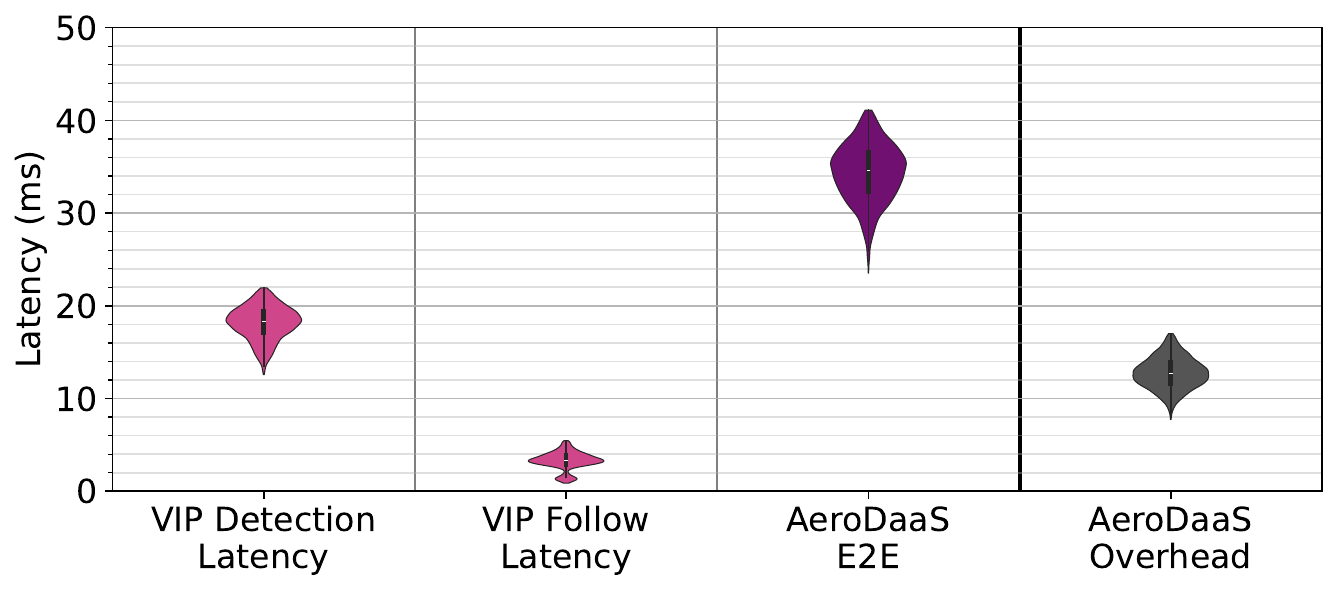}
\caption{\adp overhead and latency of analytics during simulation of sensor-data driven pattern}
\label{fig:sim_scene3_violin}
\end{figure}

\begin{figure}[t]
\centering
\includegraphics[width=0.7\columnwidth]{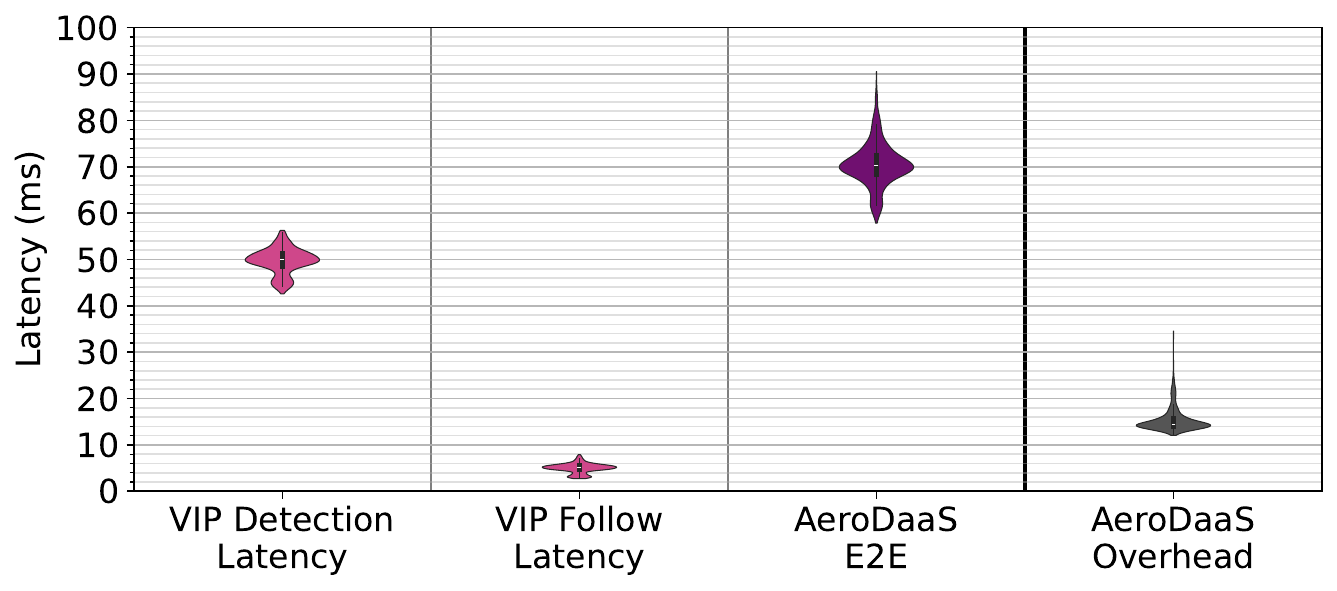}
\caption{\adp overhead and latency of analytics during real world experiment of sensor-data driven pattern}
\label{fig:real_scene3_violin}
\end{figure}

\begin{figure}[t]
\centering
\includegraphics[width=0.9\columnwidth]{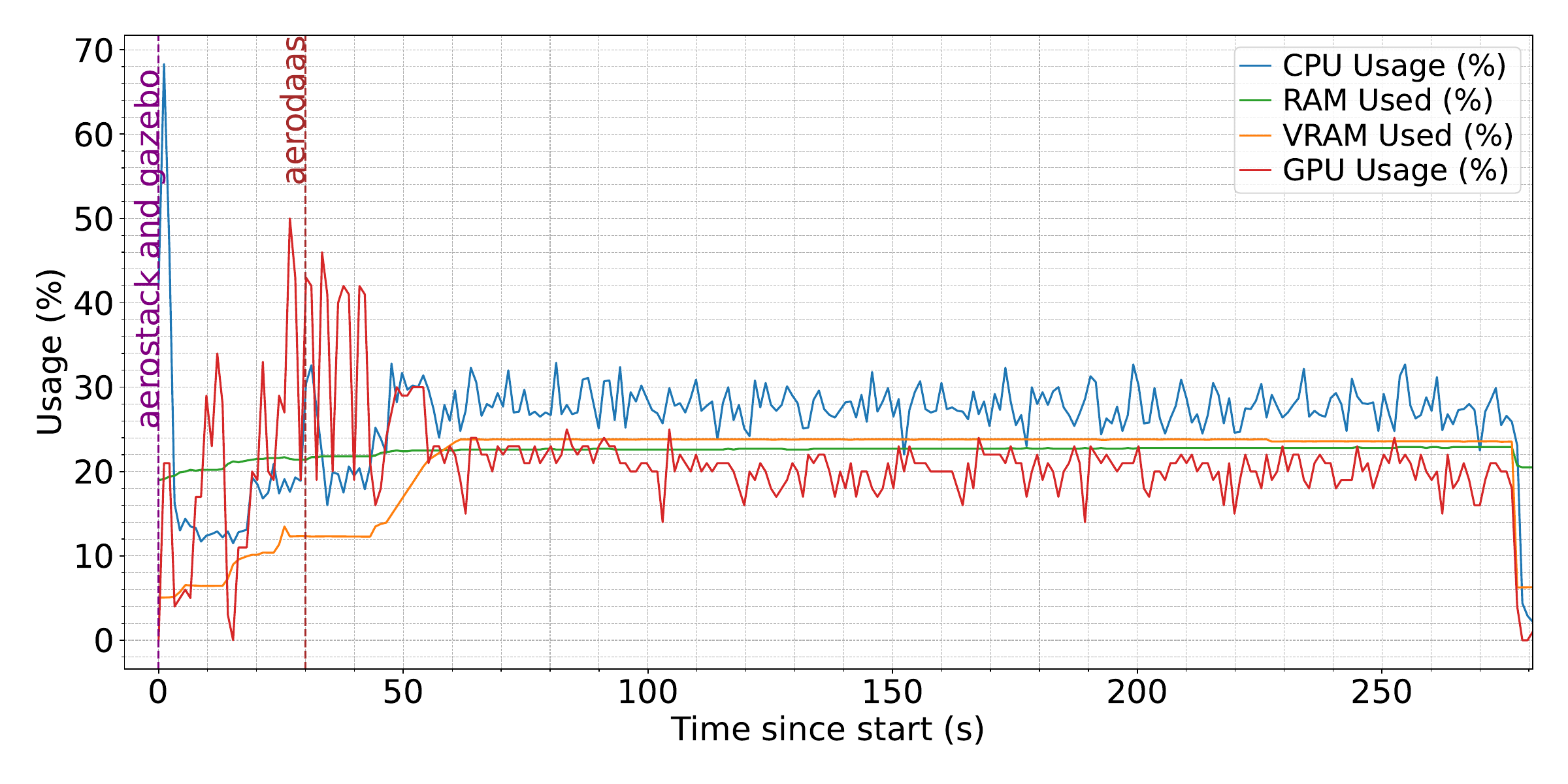}
\caption{System metrics during simulation of sensor-data driven pattern}
\label{fig:sim_scene3_vitals}
\end{figure}

\subsubsection{Overheads of \adp framework}

\begin{figure}[t]
\centering
\includegraphics[width=0.9\columnwidth]{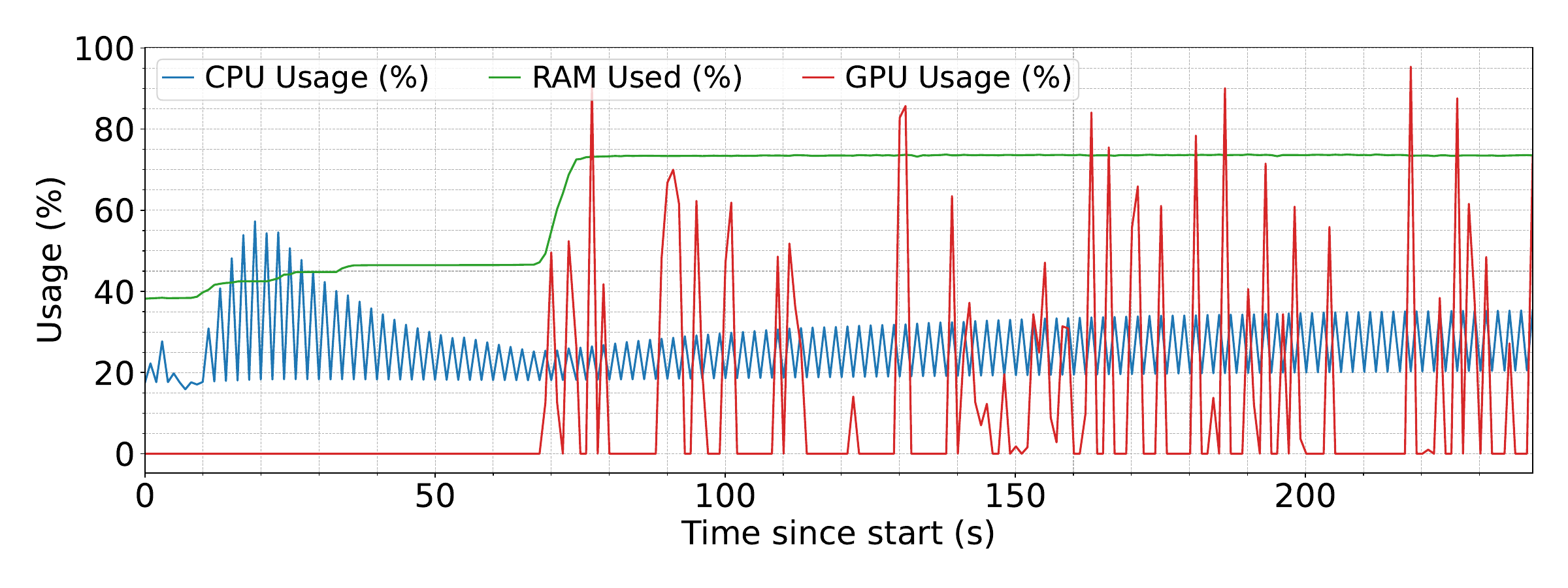}
\caption{System metrics during real world experiment of sensor-data driven pattern}
\label{fig:real_scene3_vitals}
\end{figure}

We evaluate the overhead latency of \adp using a combination of simulated and real-world analytics-driven experiments. The end-to-end latency consists receiving image frames from the drone over WiFi, executing DNN inference on the edge device, processing the inference outputs, generating navigation velocity or waypoint commands via a PID control loop, scheduling these commands, and transmitting them back to the drone. For efficient resource utilization, the DNN analytics pipeline operates on every alternate frame ($15$ FPS) in simulation and on every sixth frame ($5$ FPS) in hardware-based experiments. The \adp overhead is defined as the end-to-end latency excluding the inference and post-processing latencies.
Based on our analysis (Fig.~\ref{fig:sim_scene3_violin},~\ref{fig:real_scene3_violin}), the median end-to-end latency for simulated experiments was $35$ ms, while the hardware experiments exhibited a slightly higher latency of $73$ ms. The obsereved median latency overhead of \adp was $13$ ms in simulation and $17$ ms in hardware experiments, demonstrating that \adp introduces only a marginal computational cost while maintaining real-time responsiveness. The consistency across both experimental settings highlights the scalability and robustness of the framework for diverse mission scenarios.
Additionally, to demonstrate the deployment feasibility of \adp, we analyze its system resource consumption using a set of simulated and real-world experiments. An overview of resource utilization during the simulation experiments is shown in Fig.~\ref{fig:system-usage-sim}, which illustrates that RAM usage varied between $15\%$ and $23\%$, while VRAM usage ranged from $6\%$ to $26\%$ across different mission scenarios.
Fig.~\ref{fig:sim_scene3_vitals} and ~\ref{fig:real_scene3_vitals} show the detailed resource consumption for simulation and real-world experiments. In simulation, with \adp startup at $t=30$s and analytics activation at $t=44$s, we observe that running on a GPU workstation alongside the Gazebo simulator increased GPU memory usage by about $20\%$ ($\approx4.8$ GB on an NVIDIA RTX 3090 GPU). For the real-world experiment, the startup occurred at $t=20$s and analytics activation at $t=70$s. Running models on the edge device increased memory usage by around $40\%$ ($\approx3.2$ GB on the Orin Nano).
From Fig.~\ref{fig:sim_scene3_vitals}, it can be observed that the RAM consumption increases by $\leq1\%$ ($\approx 1$ GB) from the initialization of \adp at $t=30$s to the start of analytics at $t=44$s, indicating that \adp introduces minimal memory overhead. Overall, \adp adds $\leq0.5$ GB, i.e., only $8\%$ overhead. This trend is consistent across both GPU workstations and edge devices such as the Orin Nano, reaffirming its suitability for real-time applications.

\subsubsection{Capability of plugging in schedulers}

\begin{figure}[t]
\centering
\includegraphics[width=0.7\columnwidth]{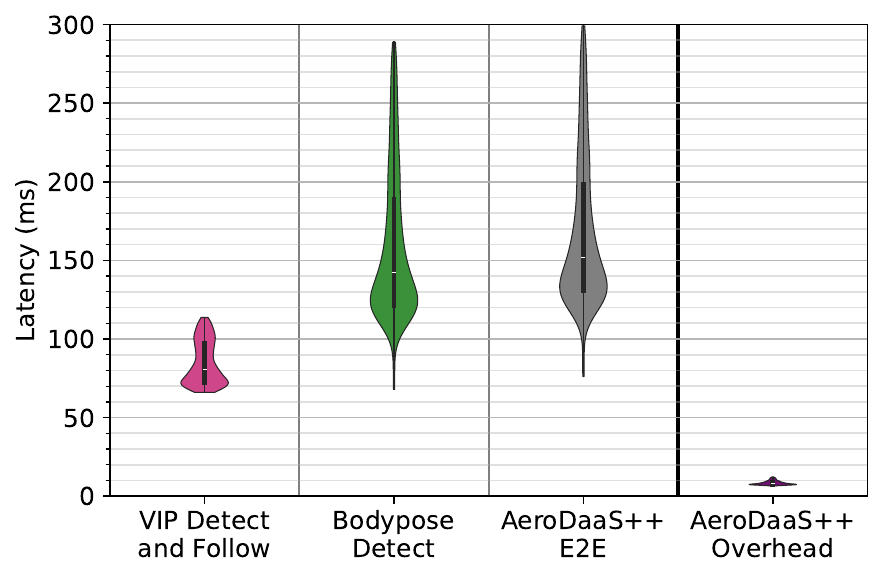}
\caption{\adp overhead and analytics scheduler latency Situation Awareness application.}
\label{fig:ocularone-violin}
\end{figure}

To demonstrate the flexibility of our system, we employed a \textit{NearestNeighbor} waypoint scheduler as a plug-in module for evaluation. This scheduler dynamically selects the next waypoint based on the nearest distance, thereby optimizing the drone's trajectory. In the Disaster Survey application, this scheduling strategy resulted in a reduction of 22 meters (refer Fig.~\ref{fig:sim_scene2_trajectory}) in total travel distance, highlighting its potential for significant energy and time savings. The scheduler's queuing capability further enables seamless switching between different missions, such as transitioning from a surveillance task to vehicle tracking, without requiring a full system reset. Similarly, the analytics scheduler coordinates onboard DNN inference tasks in parallel with navigation, ensuring efficient utilization of compute resources. We evaluated the analytics scheduler plug-in capability of \adp using the scheduler of Ocularone \cite{RAJ2025107874} with their DEMS algorithm. Fig.~\ref{fig:ocularone-violin} 
shows that the median end-to-end latency of the Situation Awareness application, which involves analytical tasks such as VIP detection and bodypose identification of the VIP, is $158$ ms with an \adp overhead of only $10$ ms. This overhead primarily consists of passing the input camera feed to the analytical models, scheduling their generated navigation commands, and transmitting those commands to the drone. The modular nature of these schedulers allows different strategies to be plugged in based on mission context, environment, and computational constraints.

\subsubsection{Concise Composition of \adp applications}

\begin{figure}[t]
\centering
\includegraphics[width=0.6\columnwidth]{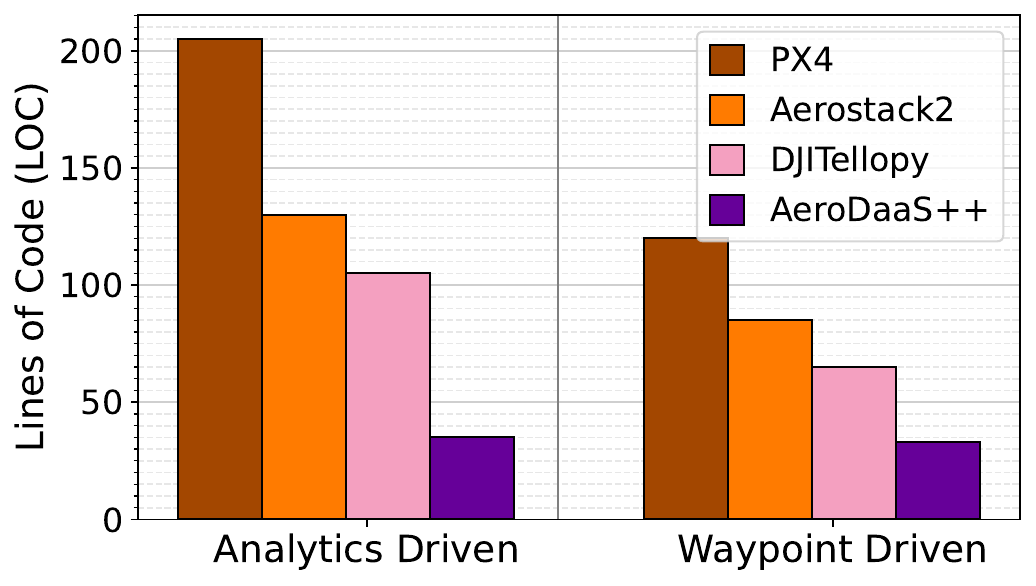}
\caption{Lines of Code (LoC) used by different frameworks}

\label{fig:loc-comparison}
\end{figure}

\adp significantly reduces the Lines of Code (LoC), by upto $5\times$, required for users to compose UAV applications compared to other platforms, making it highly efficient and user-friendly. As shown in Fig.~\ref{fig:loc-comparison}, PX4 requires around $200$ LoC for analytics-driven applications and over $130$ LoC for waypoint-driven applications, making it the most code-intensive. Aerostack2 reduces this overhead but still requires over $140$ LoC for analytics-driven tasks and more than $100$ LoC for waypoint-driven applications. Similarly, native DJI TelloPy code demands over $100$ LoC and around $70$ LoC for the two kind of tasks, respectively. In contrast, \adp requires fewer than $40$ LoC for both types of applications, offering the lowest complexity while maintaining flexibility and hardware-agnostic deployment. 

\subsubsection{Extensible analytics in \adp}

\begin{figure}[t]

\centering
\includegraphics[width=0.7\columnwidth]{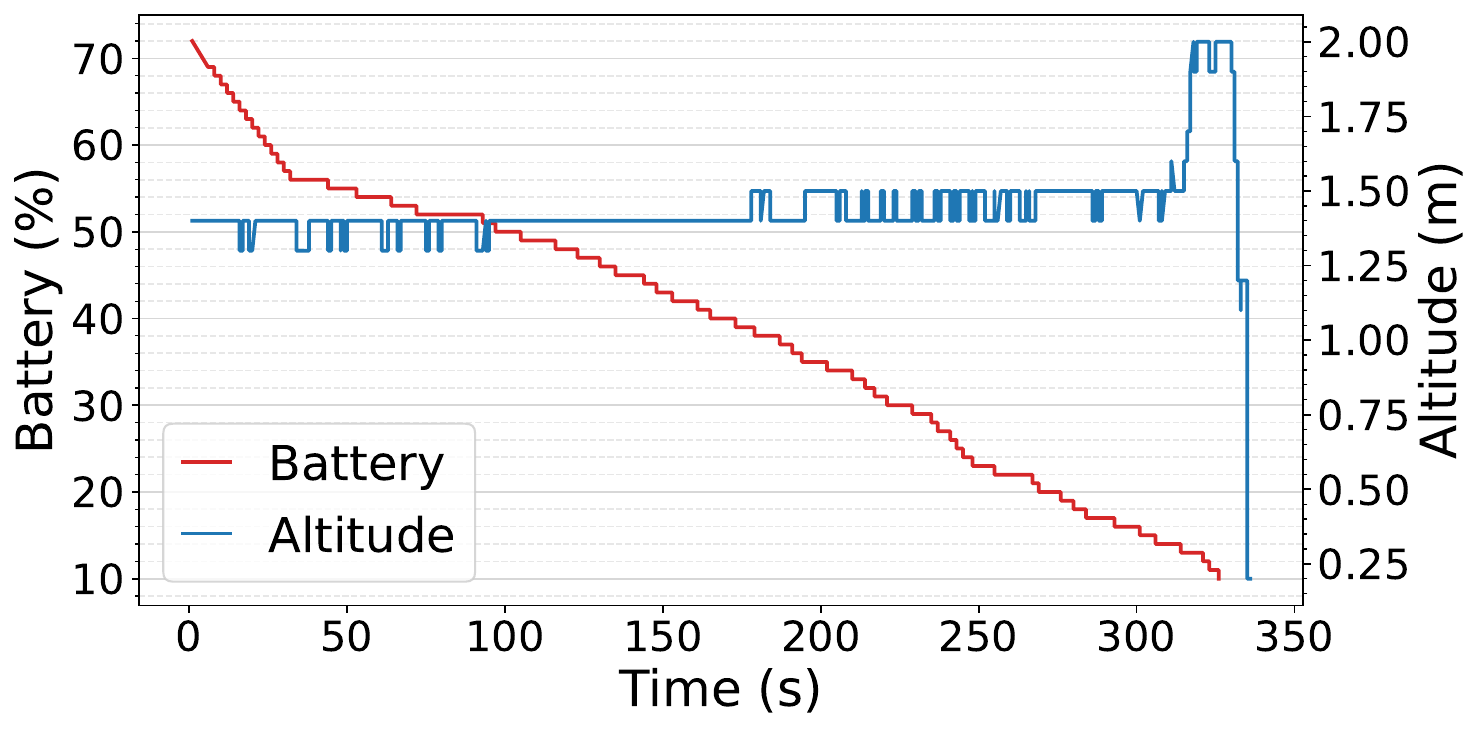}
\caption{Battery and altitude of drone during Situation Awareness application}
\label{fig:battery-monitoring}
\end{figure}

We highlight the benefits of the \texttt{MonitoringAnalytics} service, which provides access to odometry data that can be leveraged for downstream applications. For instance, Fig.~\ref{fig:battery-monitoring} illustrates the consumption of the Tello drone’s battery and altitude over time as it follows the VIP in real-world during execution of Situation Awareness application. This information can be integrated into energy-aware drone handoff algorithms to optimize mission continuity.
The PID controller employed in the Situation Awareness application was programmed to keep the drone at an altitude of $1.5$m for robust tracking of VIP but we observe that the actual altitude oscillates between $1.3$m and $1.6$m. This real-time data can be fed into the PID controller to dynamically correct errors, ensuring robust tracking performance.
Furthermore, such telemetry data can be used to analyze controller stability, detect sensor drift, and assess environmental influences. It can also log operational statistics which can later be used for predictive decision making and performance benchmarking across different flight scenarios.\\
Additionally, we demonstrate the capabilities of \texttt{Monitoring Analytics} through Farm Survey application using Gazebo. The Quadrotor Base Model drone (highlighted in blue in Fig.~\ref{fig:farm-survey-drone}) surveying a rectangular area of $10m \times 10m$ at an altitude of $1.5$m. By exposing odometry readings, we visualize the drone's trajectory in Fig.~\ref{fig:sim_scene1_trajectory}, where it takes off from the origin (0, 0, 0), follows a predefined set of waypoints covering a distance of $40$m and returns to its starting position upon completing the survey.

\section{Conclusions and Future Work}
\label{sec:conclusions}

In this paper, we present the design for \adp, a programming framework that enables intuitive design, orchestration and management of Drone-as-a-Service application across the edge-cloud continuum. It abstracts low-level drone navigation, sensing and communication complexities through high-level APIs to design drone applications across diverse drone hardware. It also integrates these with advanced user-defined analytics that can execute on edge and cloud resources. It further integrates modular pluggable schedulers for both waypoint and analytics, enabling adaptable trajectory, mission switch and efficient execution in different scenarios.
We thoroughly validated the \adp across six complex real-world DaaS applications executed using its APIs and runtime, both on real and simulated drones. The results showed that \adp efficiently handle diverse operational scenarios while maintaining minimal runtime and memory overheads on both GPU workstations and embedded edge devices confirming that \adp eases development with minimal runtime overheads.

As future work, we plan to expand the \adp APIs to support automatic high-level application code generation and controls for drone using capabilities of agentic AI. We will focus on further enhancing the scalability, resilience, and performance of \adp runtime while enabling multiple concurrent applications from different users on the same drone with coordinated execution. We also aim to introduce primitives for fleet-level operations and evaluate \adp on additional drone platforms.

\section*{Acknowledgments}
We thank Rajdeep Singh for his contributions to the development of the earlier version of AeroDaaS. We also thanks the students of the DREAM:Lab in their assistance with the drone experiments.

\bibliographystyle{unsrtnat}
\bibliography{paper}


\end{document}